\documentclass[aps,twocolumn,showpacs]{revtex4}
\usepackage{graphicx}
\include{amssym}

\begin{document}

\title{Effective multi-quark interactions with explicit breaking of chiral
       symmetry}

\author{A. A. Osipov\footnote{Email address: osipov@nu.jinr.ru},
        B. Hiller\footnote{Email address: brigitte@teor.fis.uc.pt}
    and A. H. Blin\footnote{Email address: alex@teor.fis.uc.pt}}
\affiliation{Centro de F\'{\i}sica Computacional, Departamento de
         F\'{\i}sica da Universidade de Coimbra, 3004-516 Coimbra,
         Portugal}

\begin{abstract}
In a long distance Lagrangian approach to the low lying meson phenomenology we
present and discuss the most general spin zero multi-quark interaction vertices
of non-derivative type which include a set of effective interactions
proportional to the current quark masses, breaking explicitely the chiral
$SU(3)_L\times SU(3)_R$ and $U_A(1)$ symmetries. These vertices are of the same
order in $N_c$ counting as the 't Hooft flavor determinant interaction and the
eight quark interactions which extend the original leading in $N_c$ four quark
interaction Lagrangian of Nambu and Jona-Lasinio. The $N_c$ assignements match
the counting rules based on arguments set by the scale of spontaneous chiral
symmetry breaking. With path integral bosonization techniques which take
appropriately into account the quark mass differences we derive the mesonic
Lagrangian up to three-point mesonic vertices. We demonstrate that explicit
symmetry breaking effects in interactions are essential to obtain the correct
empirical ordering and magnitude of the splitting of certain states such as
$m_K < m_\eta$ for the pseudoscalars and $m_{\kappa_0} < m_{a_0}\sim m_{f_0}$ in
the scalar sector, and achieve total agreement with the empirical low lying
meson mass spectra. With all parameters of the model fixed by the spectra we
analyze further a bulk of two body decays at tree level of the bosonic
Lagrangian: the strong decays of the scalars $\sigma\rightarrow\pi\pi$,
$f_0(980)\rightarrow\pi\pi$, $\kappa(800)\rightarrow\pi K$, $a_0(980)
\rightarrow\pi\eta$, as well as the two photon decays of $a_0(980)$, $f_0(980)$
and $\sigma$ mesons and the anomalous decays of the pseudoscalars $\pi
\rightarrow\gamma\gamma$, $\eta\rightarrow\gamma\gamma$ and $\eta'\rightarrow
\gamma\gamma$. Our results for the strong decays are within the current
expectations and the pseudoscalar radiative decays are in very good agreement
with data. The radiative decays of the scalars are smaller than the observed
values for the $f_0(980)$ and the $\sigma$, but reasonable for the $a_0$. A
detailed discussion accompanies all the results.
\end{abstract}

\pacs{11.30.Rd, 11.30.Qc, 12.39.Fe, 12.40.Yx, 14.40.Aq, 14.65.Bt}
\maketitle

\section{Introduction}

A long history of applying the Nambu -- Jona-Lasinio (NJL) model in hadron
physics shows the importance of the concept of effective multi-quark
interactions for modelling QCD at low energies. Originally formulated in terms
of the $\gamma_5$ gauge invariant nonlinear four-fermion coupling \cite{Nambu:1961,Vaks}, the model has been extended to the realistic three flavor and color
case with $U(1)_A$ breaking six-quark 't Hooft interactions \cite{Hooft:1976,Hooft:1978,Bernard:1988,Bernard:1988a,Reinhardt:1988,Weise:1990,Vogl:1990,Weise:1991,Takizawa:1990,Klevansky:1992,Hatsuda:1994,Bernard:1993,Dmitrasinovic:1990,Birse:1996,Naito:2003} and an appropriate set of eight-quark interactions \cite{Osipov:2005b}. The last ones complete the number of vertices which are important in
four dimensions for dynamical $SU(3)_L\times SU(3)_R$ chiral symmetry breaking
\cite{Andrianov:1993a,Andrianov:1993b}.

The explicit breaking of chiral symmetry in the NJL model is described by
introducing the standard light quark mass term of the QCD Lagrangian (light
means consisting of $u,d$ and $s$ quarks), e.g. \cite{Ebert:1986,Bijnens:1993}.
The current quark mass dependence is of importance for several reasons, in
particular for the phenomenological description of meson spectra and
meson-meson interactions, and for the critical point search in hot and dense
hadronic matter, where it has a strong impact on the phase diagram
\cite{Kunihiro:2010}. The values of the current quark masses are determined in
the Higgs sector of the Standard Model. In this regard they are foreign to QCD
and, at an effective description, can be included through the external sources,
interacting with the originally massless quark fields. This is why the explicit
chiral symmetry breaking (ChSB) by the standard mass term of the free
Lagrangian is only a part of the more complicated picture arising in effective
models beyond leading order \cite{Gasser:1982}. Chiral perturbation theory
\cite{Weinberg:1979,Pagels:1975,Gasser:1984,Gasser:1985} gives a well-known
example of a self consistent accounting of the mass terms, order by order, in
an expansion in the masses themselves. In fact, extended NJL-type models should
not be an exception from this rule. If one considers multi-quark effective
vertices, to the extent that $1/N_c$ suppressed 't Hooft and eight-quark terms
are included in the Lagrangian, certain mass dependent multi-quark interactions
must be also taken into account.

The aim of the present work is precisely to analyze these higher order terms in
the quark mass expansion. Our consideration proceeds along the following steps.
We start from the three-flavor NJL-type model with self-interacting massless
quarks. The $SU(3)_L\times SU(3)_R$ chiral symmetry of the Lagrangian is known
to be dynamically broken to its $SU(3)_V$ subgroup at some scale $\Lambda$,
with $\Lambda$ being one of the model parameters. There is also explicit
symmetry breaking due to the bare quark masses $\chi$, which are taken to
transform as $\chi =(3,3^*)$ under $SU(3)_L\times SU(3)_R$. Since the
Lagrangian contains, in general, an unlimited number of non-renormalizable
multi-quark and $\chi$-quark interactions (scaled by some powers of $\Lambda$),
we formulate the power counting rules to classify these vertices in accordance
with their importance for dynamical symmetry breaking. Then we bosonize the
theory by using the path-integral method. The functional integrals are
calculated in the stationary phase approximation and by using the heat kernel
technique. As a result one obtains the low-energy meson Lagrangian. At last we
fix the parameters of the model by confronting it to the experimental data. In
particular, we show the ability of the model to describe the spectrum of the
pseudo Goldstone bosons, including the fine tuning of the $\eta\!-\!\eta'$
splitting, and the spectrum of the light scalar mesons: $\sigma$ or $f_0(500)$,
$\kappa (800)$, $f_0(980)$, and $a_0(980)$.

The coupling constants of multi-quark vertices, fixed from mass-spectra, enter
the expressions for meson decay amplitudes and lead to a bulk of model
predictions. It is interesting to note that certain multi-quark vertices of the
model encode implicitly in the couplings of the tree level bosonized Lagrangian
the signature of $q\bar q$ and more complex quark structures which are
elsewhere obtained by considering explicitly meson loop corrections, tetraquark
configurations and so on \cite{Jaffe:1977,Black:1999,Wong:1980,Narrison:1986,Beveren:1986, Latorre:1985,Alford:1998,Achasov:1984,Isgur:1990,Schechter:2008,Schechter:2009,Close:2002,Klempt:2007}. It seems appropriate, therefore, to examine
the possible physics opportunities connected with the discovery and study of
such multi-quark structures in hadrons. For instance, by calculating the mass
spectra and the strong decays of the scalars, one can realize which multi-quark
interactions are most relevant at the scale of spontaneous ChSB. On the other
hand, by analyzing the two photon radiative decays, where a different scale,
associated with the electromagnetic interaction, comes into play, one can study
the possible recombinations of quarks inside the hadron. We will show, for
example, that the $a_0(980)$ meson couples with a large strength of the
multi-quark components to the two kaon channel in its strong decay to two
pions, but evidences a dominant $q\bar q$ component in its radiative decay. As
opposed to this, the $\sigma$ and $f_0(980)$ mesons do not display an enhanced
$q\bar q$ component neither in their two photon decays nor in the strong
decays.

There are several direct motivations for this work. In the first place, the
quark masses are the only parameters of the QCD Lagrangian which are
responsible for the explicit ChSB, and it is important for the effective theory
to trace this dependence in full detail. In this paper it will be argued that
it is from the point of view of the $1/N_c$ expansion that the new quark mass
dependent interactions must be included in the NJL-type Lagrangian already when
the $U(1)_A$ breaking 't Hooft determinantal interaction is considered. This
important point is somehow completely ignored in the current literature.

A second reason is that nowadays it is getting clear that the eight-quark
interactions, which are almost inessential for the mesonic spectra in the
vacuum, can be important for the quark matter in a strong magnetic background
\cite{Hiller:2007,Gatto:2010,Gatto:2011,Frasca:2011,Gatto:2012}. The simplest
next possibility is to add to that picture a set of new effective
quark-mass-dependent interactions, discussed in this work. Such feature of the
quark matter has not been studied yet, but probably contains interesting
physics.

Further motivation comes from the hadronic matter studies in a hot and dense
environment. It is known that lattice QCD at finite density suffers from
the numerical sign problem. This is why the phase diagram is notoriously
difficult to compute ``ab initio'', except for the extremely high density regime
where perturbative QCD methods are applicable. In such circumstances effective
models designed to shed light on the phase structure of QCD are valuable,
especially if such models are known to be successful in the description of the
hadronic matter at zero temperature and density. Reasonable modifications of
the NJL model are of special interest in this context and our work aims also at
future applications in that area.

The paper is organized as follows. In section II the effective Lagrangian in
terms of quark degrees of freedom and bosonic sources with specific quantum
numbers is derived using a classification scheme which selects all possible
non-derivative vertices according to the symmetries of the strong interaction
and which are relevant at the scale $\Lambda$ of spontaneous chiral symmetry
breaking. It is then shown that this scheme can be equally organized in terms
of the large $N_c$ counting rules, which in turn allow to attribute to the
couplings of the interactions encoded signatures of $q\bar q$ and more complex
structures involving four fermions. We obtain in this section also that a set
of interactions lead to the Lagrangian specific Kaplan-Manohar ambiguity
associated with the current quark masses.

In section III we proceed to bosonize the multi-quark Lagrangian in two steps.
First, we introduce in section III-A a set of auxiliary scalar fields. By
these new variables the multi-quark interactions can be brought to the Yukawa
form that is quadratic in Fermi fields. Consequently one obtains a Gauss-type
integral over quarks, and a set of integrals over auxiliary fields. The
latter are evaluated by the stationary phase method. We obtain here the
vertices up to the cubic power in the meson fields, needed for the study of the
meson spectra and of the two-body decays. Then, in section III-B, we integrate
over quark fields. The arising quark determinant of the Dirac operator is a
complicated non-local functional of the collective meson fields. We calculate
it in the low-energy regime by using the Schwinger-DeWitt technique, based on
the heat kernel expansion. In this approximation one can adequately incorporate
the effect of different quark masses contained in the modulus of the one-loop
quark determinant. We derive the kinetic terms of the collective meson fields,
as well as the heat kernel part of contributions to meson masses and
interactions. In the end of this section we present the complete bosonized
Lagrangian, give the mixing angle conventions used, and the expressions for
the strong decay widths. In section III-C we obtain the expressions for the
radiative widths of the pseudoscalars and scalars.

In section IV we present the numerical results and discussion, in IV-A for the
meson mass spectra and weak decay constants, in IV-B for the strong decays and
in IV-C for the radiative decays.

We conclude in section V with a summary of the main results.

\section{Effective multi-quark interactions}

The chiral quark Lagrangian has predictive power for the energy range which is
of order $\Lambda\simeq 4\pi f_\pi\sim 1$\ GeV \cite{Georgi:1984}. $\Lambda$
characterizes the spontaneous chiral symmetry breaking scale. Consequently, the
effective multi-quark interactions, responsible for this dynamical effect, are
suppressed by $\Lambda$, which provides a natural expansion parameter in the
chiral effective Lagrangian. The scale above which these interactions disappear
and QCD becomes perturbative enters the NJL model as an ultraviolet cut-off for
the quark loops. Thus, to build the NJL type Lagrangian we have only three
elements: the quark fields $q$, the scale $\Lambda$, and the external sources
$\chi$, which generate explicit symmetry breaking effects -- resulting in mass
terms and mass-dependent interactions.

The color quark fields possess definite transformation properties with respect
to the chiral flavor $U(3)_L\times U(3)_R$ global symmetry of the QCD
Lagrangian with three massless quarks (in the large $N_c$ limit). It is
convenient to introduce the $U(3)$ Lie-algebra valued field $\Sigma =
\frac{1}{2}(s_a-ip_a)\lambda_a$, where $s_a=\bar q\lambda_aq$, $p_a=\bar
q\lambda_ai\gamma_5q$, and $a=0,1,\ldots ,8$, $\lambda_0=\sqrt{2/3}\times 1$,
$\lambda_a$ being the standard $SU(3)$ Gell-Mann matrices for $1\leq a \leq 8$.
Under chiral transformations: $q'=V_Rq_R+V_Lq_L$, where $q_R=P_R q, q_L=P_Lq$,
and $P_{R,L}=\frac{1}{2}(1\pm\gamma_5)$. Hence, $\Sigma'=V_R\Sigma V_L^\dagger$,
and $\Sigma^{\dagger'}=V_L\Sigma^\dagger V_R^\dagger$. The transformation property
of the source is supposed to be $\chi' =V_R\chi V_L^\dagger$.

Any term of the effective multi-quark Lagrangian without derivatives can be
written as a certain combination of fields which is invariant under chiral
$SU(3)_R\times SU(3)_L$ transformations and conserves $C, P$ and $T$ discrete
symmetries. These terms have the general form
\begin{equation}
\label{genL}
   L_i\sim \frac{\bar g_i}{\Lambda^\gamma}\chi^\alpha\Sigma^\beta,
\end{equation}
where $\bar g_i$ are dimensionless coupling constants (starting from eq.
(\ref{h}) the dimensional couplings $g_i=\bar g_i/\Lambda^{\gamma}$ will be also
considered). Using dimensional arguments we find (in four dimensions)
$\alpha+3\beta -\gamma =4$, with integer values for $\alpha, \beta$ and
$\gamma$.

We obtain a second restriction by considering only the vertices which make
essential contributions to the gap equations in the regime of dynamical chiral
symmetry breaking, i.e. we collect only the terms whose contributions to the
effective potential survive at $\Lambda\to\infty$. We get this information by
contracting quark lines in $L_i$, finding that this term contributes to the
power counting of $\Lambda$ in the effective potential as $\sim
\Lambda^{2\beta-\gamma}$, i.e. we obtain that $2\beta-\gamma \geq 0$ (we used
the fact that in four dimensions each quark loop contributes as $\Lambda^2$).

Combining both restrictions we come to the conclusion that only vertices with
\begin{equation}
\label{ineq}
   \alpha +\beta \leq 4
\end{equation}
must be taken into account in the approximation considered. On the basis of
this inequality one can conclude that (i) there are only four classes of
vertices which contribute at $\alpha=0$; those are four, six and eight-quark
interactions, corresponding to $\beta=2,3$ and $4$ respectively; the $\beta=1$
class is forbidden by chiral symmetry requirements; (ii) there are only six
classes of vertices depending on external sources $\chi$, they are: $\alpha
=1, \beta =1,2,3$; $\alpha =2, \beta =1,2$; and $\alpha =3, \beta =1$.

Let us consider now the structure of multi-quark vertices in detail
\cite{Osipov:2013}. The Lagrangian corresponding to the case (i) is well known
\begin{eqnarray}
\label{L-int}
   L_{int}&=&\frac{\bar G}{\Lambda^2}\mbox{tr}\left(\Sigma^\dagger\Sigma\right)
   +\frac{\bar\kappa}{\Lambda^5}\left(\det\Sigma+\det\Sigma^\dagger\right)
   \nonumber \\
   &+&\frac{\bar g_1}{\Lambda^8}\left(\mbox{tr}\,\Sigma^\dagger\Sigma\right)^2
   +\frac{\bar g_2}{\Lambda^8}\mbox{tr}
   \left(\Sigma^\dagger\Sigma\Sigma^\dagger\Sigma\right).
\end{eqnarray}
It contains four dimensionful couplings $G, \kappa, g_1, g_2$.

The second group (ii) contains eleven terms
\begin{equation}
   L_\chi =\sum_{i=0}^{10}L_i,
\end{equation}
where
\begin{eqnarray}
\label{L-chi-1}
   L_0&=&-\mbox{tr}\left(\Sigma^\dagger\chi +\chi^\dagger\Sigma\right)
   \nonumber \\
   L_1&=&-\frac{\bar\kappa_1}{\Lambda}e_{ijk}e_{mnl}
   \Sigma_{im}\chi_{jn}\chi_{kl}+h.c.
   \nonumber \\
   L_2&=&\frac{\bar\kappa_2}{\Lambda^3}e_{ijk}e_{mnl}
   \chi_{im}\Sigma_{jn}\Sigma_{kl}+h.c.
   \nonumber \\
   L_3&=&\frac{\bar g_3}{\Lambda^6}\mbox{tr}
   \left(\Sigma^\dagger\Sigma\Sigma^\dagger\chi\right)+h.c.
   \nonumber \\
   L_4&=&\frac{\bar g_4}{\Lambda^6}\mbox{tr}\left(\Sigma^\dagger\Sigma\right)
   \mbox{tr}\left(\Sigma^\dagger\chi\right)+h.c.
   \nonumber \\
   L_5&=&\frac{\bar g_5}{\Lambda^4}\mbox{tr}\left(\Sigma^\dagger\chi
   \Sigma^\dagger\chi\right)+h.c.
   \nonumber \\
   L_6&=&\frac{\bar g_6}{\Lambda^4}\mbox{tr}\left(\Sigma\Sigma^\dagger\chi
   \chi^\dagger +\Sigma^\dagger\Sigma\chi^\dagger\chi\right)
   \nonumber \\
   L_7&=&\frac{\bar g_7}{\Lambda^4}\left(\mbox{tr}\Sigma^\dagger\chi
   + h.c.\right)^2
   \nonumber \\
   L_8&=&\frac{\bar g_8}{\Lambda^4}\left(\mbox{tr}\Sigma^\dagger\chi
   - h.c.\right)^2
   \nonumber \\
   L_9&=&-\frac{\bar g_9}{\Lambda^2}\mbox{tr}\left(\Sigma^\dagger\chi
   \chi^\dagger\chi\right)+h.c.
   \nonumber \\
   L_{10}&=&-\frac{\bar g_{10}}{\Lambda^2}\mbox{tr}\left(\chi^\dagger\chi\right)
   \mbox{tr}\left(\chi^\dagger\Sigma\right)+h.c.
\end{eqnarray}
Each term in the Lagrangian $L_6$ is hermitian by itself, but because of the
parity symmetry of strong interactions, which transforms one of them into the
other, they have a common coupling $\bar g_6$.

Some useful insight into the Lagrangian above can be obtained by considering it
from the point of view of the $1/N_c$ expansion. Indeed, the number of color
components of the quark field $q^i$ is $N_c$, hence summing over color indices
in $\Sigma$ gives a factor of $N_c$, i.e. one counts $\Sigma\sim N_c$.

The cut-off $\Lambda$ that gives the right dimensionality to the multi-quark
vertices scales as $\Lambda\sim N_c^0=1$, as a direct consequence of the gap
equations (see eq. (\ref{gap}) below), which imply $1\sim N_c G \Lambda^2$; on
the other hand, since the leading quark contribution to the vacuum energy is
known to be of order $N_c$, the first term in (\ref{L-int}) is estimated as
$N_c$, and we conclude that $G\sim 1/N_c$.

Furthermore, the $U(1)_A$ anomaly contribution (the second term in
(\ref{L-int})) is suppressed by one power of $1/N_c$, it yields $\kappa\sim
1/N_c^3$.

The last two terms in (\ref{L-int}) have the same $N_c$ counting as the 't
Hooft term. They are of order $1$. Indeed, Zweig's rule violating effects
are always of order $1/N_c$ with respect to the leading order contribution
$\sim N_c$. This reasoning helps us to find $g_1\sim 1/N_c^4$. The term with
$g_2\sim 1/N_c^4$ is also $1/N_c$ suppressed. It represents the next to the
leading order contribution with one internal quark loop in $N_c$ counting.
Such vertex contains the admixture of the four-quark component $\bar qq\bar qq$
to the leading quark-antiquark structure at $N_c\to\infty$.

Next, all terms in eq. (\ref{L-chi-1}), except $L_0$, are of order 1. The
argument is just the same as before: this part of the Lagrangian is obtained
by succesive insertions of the $\chi$-field ($\chi$ counts as $\chi\sim 1$) in
place of $\Sigma$ fields in the already known $1/N_c$ suppressed vertices. It
means that $\kappa_1, g_9, g_{10}\sim 1/N_c$, $\kappa_2, g_5, g_6, g_7,
g_8\sim 1/N_c^2$, and $g_3, g_4\sim 1/N_c^3$.

There are two important conclusions here. The first is that at leading order
in $1/N_c$ only two terms contribute: the first term of eq. (\ref{L-int}), and
the first term of eq. (\ref{L-chi-1}). This corresponds exactly to the standard
NJL model picture, where mesons are pure $\bar qq$ states with constituents
which have a non-zero bare mass. At the next to leading order we have
thirteen terms additionally. They trace the Zweig's rule violating effects
$(\kappa, \kappa_1, \kappa_2, g_1, g_4, g_7, g_8, g_{10})$, and an admixture of
the four-quark component to the $\bar qq$ one ($g_2, g_3, g_5, g_6$, $g_9$).
Only the phenomenology of the last three terms from eq. (\ref{L-int}) has been
studied until now. We must still understand the role of the other ten terms to
be consistent with the generic $1/N_c$ expansion of QCD.

The second conclusion is that the $N_c$ counting justifies the classification
of the vertices made above on the basis of the inequality (\ref{ineq}). This is
seen as follows: the equivalent inequality $\lceil (\alpha +\beta )/2\rceil
\leq 2$ is obtained by restricting the multi-quark Lagrangian to terms that do
not vanish at $N_c\to\infty$ (it follows from (\ref{genL}) that $\beta -\lceil
\gamma /2\rceil\geq 0$ by noting that $\bar g_i\sim 1/N_c^{\lceil\gamma /2\rceil}$,
where $\lceil\gamma /2\rceil$ is the nearest integer greater than or equal to
$\gamma /2$).

The total Lagrangian is the sum
\begin{equation}
\label{LQ}
   L=\bar qi\gamma^\mu\partial_\mu q+L_{int}+L_\chi.
\end{equation}
In this $SU(3)_L\times SU(3)_R$ symmetric chiral Lagrangian we neglect terms
with derivatives in the multi-quark interactions, as usually assumed in the
NJL model. We follow this approximation, because the specific questions for
which these terms might be important, e.g. the radial meson excitations, or
the existence of some inhomogeneous phases, characterized by a spatially
varying order parameter, are not the goal of this work.

Finally, having all the building blocks conform with the symmetry pattern of
the model, one is now free to choose the external source $\chi$. Putting $\chi
={\cal M}/2$, where
$$
   {\cal M}=\mbox{diag}(\mu_u, \mu_d, \mu_s),
$$
we obtain a consistent set of explicitly breaking chiral symmetry terms. This
leads to the following mass dependent part of the NJL Lagrangian
\begin{equation}
   L_\chi \to L_\mu= -\bar qmq+\sum_{i=2}^8L_i'
\end{equation}
where the current quark mass matrix $m$ is equal to
\begin{eqnarray}
\label{cqmm}
   m&=&{\cal M}+\frac{\bar\kappa_1}{\Lambda}\left(\det {\cal M}
   \right){\cal M}^{-1}+\frac{\bar g_9}{4\Lambda^2}{\cal M}^3
   \nonumber \\
   &+&\frac{\bar g_{10}}{4\Lambda^2}\left(\mbox{tr}{\cal M}^2\right){\cal M},
\end{eqnarray}
and
\begin{equation}
\label{L-chi-2}
   \begin{array}{lcr}
   L_2'=\frac{\bar\kappa_2}{2\Lambda^3}e_{ijk}e_{mnl}
   {\cal M}_{im}\Sigma_{jn}\Sigma_{kl}+h.c.
   \\ \\
   L_3'=\frac{\bar g_3}{2\Lambda^6}\mbox{tr}
   \left(\Sigma^\dagger\Sigma\Sigma^\dagger{\cal M}\right)+h.c.
   \\ \\
   L_4'=\frac{\bar g_4}{2\Lambda^6}\mbox{tr}\left(\Sigma^\dagger\Sigma\right)
   \mbox{tr}\left(\Sigma^\dagger{\cal M}\right)+h.c.
   \\ \\
   L_5'=\frac{\bar g_5}{4\Lambda^4}\mbox{tr}\left(\Sigma^\dagger{\cal M}
   \Sigma^\dagger{\cal M}\right)+h.c.
   \\ \\
   L_6'=\frac{\bar g_6}{4\Lambda^4}\mbox{tr}\left[{\cal M}^2
   \left(\Sigma\Sigma^\dagger +\Sigma^\dagger\Sigma\right)\right]
   \\ \\
   L_7'=\frac{\bar g_7}{4\Lambda^4}\left(\mbox{tr}\Sigma^\dagger{\cal M}
   + h.c.\right)^2
   \\ \\
   L_8'=\frac{\bar g_8}{4\Lambda^4}\left(\mbox{tr}\Sigma^\dagger{\cal M}
   - h.c.\right)^2
   \end{array}
\end{equation}

Let us note that there is a definite freedom in the definition of the external
source $\chi$. In fact, the sources
\begin{eqnarray}
   \chi^{(c_i)}&=&\chi +\frac{c_1}{\Lambda}
   \left(\det\chi^\dagger\right)\chi\left(\chi^\dagger\chi\right)^{-1}+
   \frac{c_2}{\Lambda^2}\chi\chi^\dagger\chi
   \nonumber \\
   &+&\frac{c_3}{\Lambda^2}\mbox{tr}\left(\chi^\dagger\chi\right)\chi
\end{eqnarray}
with three independent constants $c_i$ have the same symmetry transformation
property as $\chi$. Therefore, we could have used $\chi^{(c_i)}$ everywhere
that we used $\chi$. As a result, we would come to the same Lagrangian with
the following redefinitions of couplings
\begin{eqnarray}
\label{rt}
   &&\bar\kappa_1\to\bar\kappa_1'=\bar\kappa_1+\frac{c_1}{2}, \quad
   \bar g_5\to\bar g_5'=\bar g_5-\bar\kappa_2c_1,
   \nonumber \\
   &&\bar g_7\to\bar g_7'=\bar g_7+\frac{\bar\kappa_2}{2}c_1,  \quad
   \bar g_8\to\bar g_8'=\bar g_8 +\frac{\bar\kappa_2}{2}c_1, \quad
   \nonumber \\
   &&\bar g_9\to\bar g_9'=\bar g_9 +c_2 -2\bar\kappa_1 c_1, \quad
   \nonumber \\
   &&\bar g_{10}\to\bar g_{10}'=\bar g_{10}+c_3+2\bar\kappa_1 c_1.
\end{eqnarray}
Since $c_i$ are arbitrary parameters, this corresponds to a continuous family
of equivalent Lagrangians. This family reflects the known Kaplan -- Manohar
ambiguity \cite{Manohar:1986,Leutwyler:1990,Donoghue:1992,Leutwyler:1996} in
the definition of the quark mass, and means that several different parameter
sets (\ref{rt}) may be used to represent the data. In particular, without loss
of generality we can use the repara\-metrization freedom to obtain the set
with $\bar\kappa_1' =\bar g_9' =\bar g_{10}'=0$.

The effective multi-quark Lagrangian can be written now as
\begin{equation}
\label{qlagr}
   L=\bar q(i\gamma^\mu\partial_\mu -m)q+L_{int}+\sum_{i=2}^8L_i'.
\end{equation}
It contains eighteen parameters: the scale $\Lambda$, three parameters which
are responsible for explicit chiral symmetry breaking $\mu_u,\mu_d,\mu_s$,
and fourteen interaction couplings $\bar G, \bar\kappa, \bar\kappa_1, \bar
\kappa_2$, $\bar g_1,\ldots,\bar g_{10}$. Three of them, $\bar\kappa_1,
\bar g_9, \bar g_{10}$, contribute to the current quark masses $m$. Seven more
describe the strength of multi-quark interactions with explicit symmetry
breaking effects. These vertices contain new details of the quark dynamics
which have not been studied yet in any NJL-type models. We shall now see how
important they are.

\section{Bosonization: meson masses and decays}

\subsection{Stationary phase contribution}

The model can be solved by path integral bosonization of the quark Lagrangian
(\ref{qlagr}). Indeed, following \cite{Reinhardt:1988} we may equivalently
introduce auxiliary fields $s_a=\bar q\lambda_aq,\, p_a=\bar qi\gamma_5
\lambda_aq$, and physical scalar and pseudoscalar fields $\sigma =\sigma_a
\lambda_a,\,\phi = \phi_a\lambda_a$. In these variables the Lagrangian is a
bilinear form in quark fields (once the replacement has been done the quarks
can be integrated out giving us the kinetic terms for the physical fields
$\phi$ and $\sigma$)
\begin{eqnarray}
\label{L}
   L&\!=\!&\bar q\left(i\gamma^\mu\partial_\mu -\sigma - i\gamma_5\phi
   \right)q + L_{aux}, \nonumber \\
   L_{aux}&\!=\!&s_a\sigma_a + p_a\phi_a - s_am_a +L_{int}(s,p) \nonumber \\
   &\!+\!&\sum_{i=2}^8L_i'(s,p,\mu).
\end{eqnarray}
It is clear, that after the elimination of the fields $\sigma,\,\phi$ by means
of their classical equations of motion, one can rewrite this Lagrangian in its
original form (\ref{qlagr}). The term bilinear in the quark fields in (\ref{L})
will be integrated out using the heat kernel technique in the next subsection.
The remaining higher order quark interactions collected in $ L_{aux}$  will be
integrated in the stationary phase approximation (SPA). In terms of auxiliary
bosonic variables one has
\begin{eqnarray}
   L_{int}(s,p)&\!=\!&L_{4q}+L_{6q}+L_{8q}^{(1)}+L_{8q}^{(2)},
   \nonumber \\
   L_{4q}(s,p)&\!=\!&\frac{\bar G}{2\Lambda^2}\left(s_a^2+p_a^2\right),
   \nonumber \\
   L_{6q}(s,p)&\!=\!&\frac{\bar\kappa}{4\Lambda^5}A_{abc}s_a(s_bs_c-3p_bp_c),
   \\
   L_{8q}^{(1)}(s,p)&\!=\!&\frac{\bar g_1}{4\Lambda^8}\left(s_a^2+p_a^2\right)^2,
   \nonumber \\
   L_{8q}^{(2)}(s,p)&\!=\!&\frac{\bar g_2}{8\Lambda^8}\left[d_{abe}d_{cde}
   \left(s_as_b+p_ap_b\right)\left(s_cs_d+p_cp_d\right)\right.
   \nonumber \\
   &\!+\!&\left.4f_{abe}f_{cde}s_as_cp_bp_d\right], \nonumber
\end{eqnarray}
and the quark mass dependent part is as follows
\begin{eqnarray}
\label{L-chi-3}
   L_2'&\!=\!&\frac{3\bar\kappa_2}{2\Lambda^3}A_{abc}\mu_a\left(s_bs_c-p_bp_c
   \right), \nonumber \\
   L_3'&\!=\!&\frac{\bar g_3}{4\Lambda^6}\mu_a
   \left[d_{abe}d_{cde}s_b\left(s_cs_d+p_cp_d\right)-2f_{abe}f_{cde}p_bp_cs_d
   \right],
   \nonumber \\
   L_4'&\!=\!&\frac{\bar g_4}{2\Lambda^6}\mu_bs_b\left(s_a^2+p_a^2\right),
   \nonumber \\
   L_5'&\!=\!&\frac{\bar g_5}{4\Lambda^4}\mu_b\mu_d\left(d_{abe}d_{cde}-
   f_{abe}f_{cde}\right)\left(s_as_c-p_ap_c\right),
   \nonumber \\
   L_6'&\!=\!&\frac{\bar g_6}{4\Lambda^4}\mu_a\mu_bd_{abe}d_{cde}
   \left(s_cs_d+p_cp_d\right),
   \nonumber \\
   L_7'&\!=\!&\frac{\bar g_7}{\Lambda^4}\left(\mu_as_a\right)^2,
   \nonumber \\
   L_8'&\!=\!&-\frac{\bar g_8}{\Lambda^4}\left(\mu_ap_a\right)^2,
\end{eqnarray}
where
\begin{equation}
   A_{abc}=\frac{1}{3!}e_{ijk}e_{mnl}(\lambda_a)_{im}(\lambda_b)_{jn}
   (\lambda_c)_{kl},
\end{equation}
and the $U(3)$ antisymmetric $f_{abc}$ and symmetric $d_{abc}$ constants are
standard.

Our final goal is to clarify the phenomenological role of the mass-dependent
terms described by the Lagrangian densites of eq. (\ref{L-chi-3}). We can gain
some understanding of this by considering the low-energy meson dynamics which
follows from our Lagrangian. For that we must exclude quark degrees of freedom
in (\ref{L}), e.g., by integrating them out from the corresponding generating
functional. The standard Gaussian path integral leads us to the fermion
determinant, which we expand by using a heat-kernel technique
\cite{Osipov:2006a,Osipov:2001,Osipov:2001a,Osipov:2001b}. The remaining part
of the Lagrangian, $L_{aux}$, depends on auxiliary fields which do not have
kinetic terms. The equations of motion of such a static system are the
extremum conditions
\begin{equation}
\label{sp}
   \frac{\partial L}{\partial s_a}=0, \quad \frac{\partial L}{\partial p_a}=0,
\end{equation}
which must be fulfilled in the neighbourhood of the uniform vacuum state of the
theory. To take this into account one should shift the scalar field $\sigma\to
\sigma +M$. The new $\sigma$-field has a vanishing vacuum expectation value
$\langle\sigma\rangle =0$, describing small amplitude fluctuations about the
vacuum, with $M$ being the mass of constituent quarks. We seek solutions of eq.
(\ref{sp}) in the form:
\begin{eqnarray}
\label{st}
   s_a^{st}&=&h_a+ h_{ab}^{(1)}\sigma_b + h_{abc}^{(1)}\sigma_b\sigma_c
   +h_{abc}^{(2)}\phi_b\phi_c + \ldots
   \nonumber \\
    p_a^{st}&=&h_{ab}^{(2)}\phi_b + h_{abc}^{(3)}\phi_b\sigma_c +\ldots
\end{eqnarray}
Eqs. (\ref{sp}) determine all coefficients of this expansion giving rise to a
system of cubic equations to obtain $h_a$, and the full set of recurrence
relations to find higher order coefficients in (\ref{st}). We can gain some
insight into the physical meaning of these parameters if we calculate the
Lagrangian density $L_{aux}$ on the stationary trajectory. In fact, using the
recurrence relations, we are led to the result
\begin{eqnarray}
\label{lam}
   L_{aux}\!\!\!\!\!\!
   &&=h_a\sigma_a+\frac{1}{2}\,h_{ab}^{(1)}\sigma_a\sigma_b
      +\frac{1}{2}\,h_{ab}^{(2)}\phi_a\phi_b  \\
   &&+\,\frac{1}{3}\,\sigma_a\left[h^{(1)}_{abc}\sigma_b\sigma_c
   +\left(h^{(2)}_{abc}+h^{(3)}_{bca}\right)\phi_b\phi_c\right]
   + \ldots \nonumber
\end{eqnarray}
Indicated are all the terms which are necessary to analyze the mass spectra
and two particle decays. Here $h_a$ define the quark condensates, $h_{ab}^{(1)}$,
$h_{ab}^{(2)}$ contribute to the masses of scalar and pseudoscalar states, and
higher order $h$'s are the couplings that measure the strength of the
meson-meson interactions. The transition from the Lagrangian $L_{aux}(s,p)$ in
(\ref{L}) to its form $L_{aux}(\sigma,\phi)$ in (\ref{lam}) can be viewed as a
Legendre transformation.

We proceed now to explain the details of determining $h$. We address first the
coefficients $h_a$, $h_{ab}^{(1)}$, and $h_{ab}^{(2)}$. In particular, eq.
(\ref{sp}) states that $h_a=0$, if $a\neq 0,3,8$, while $h_\alpha$ ($\alpha
=0,3,8$), after the convenient redefinition to the flavor indices $i=u,d,s$
\begin{equation}
   h_\alpha =e_{\alpha i} h_i, \quad
   e_{\alpha i}=\frac{1}{2\sqrt 3}\left(
          \begin{array}{ccc}
          \sqrt 2&\sqrt 2&\sqrt 2 \\
          \sqrt 3&-\sqrt 3& 0 \\
          1&1&-2
          \end{array} \right),
\end{equation}
satisfy the following system of cubic equations
\begin{eqnarray}
\label{h}
   &&\Delta_i + \frac{\kappa}{4}t_{ijk}h_jh_k +\frac{h_i}{2}\left(
     2G+g_1h^2 + g_4\mu h\right) + \frac{g_2}{2}h_i^3 \nonumber \\
   &&+\frac{\mu_i}{4}\left[3g_3h_i^2 +g_4h^2 +2(g_5+g_6)\mu_ih_i
     +4g_7 \mu h\right] \nonumber \\
   &&+\kappa_2t_{ijk}\mu_jh_k=0.
\end{eqnarray}
Here $\Delta_i=M_i-m_i$; $t_{ijk}$ is a totally symmetric quantity, whose
nonzero components are $t_{uds}=1$; there is no summation over the open index
$i$ but we sum over the dummy indices, e.g. $h^2=h_u^2+h_d^2+h_s^2,
\mu h=\mu_uh_u+\mu_dh_d+\mu_sh_s$.

In particular, eq. (\ref{cqmm}) reads in this basis
\begin{equation}
\label{cqm-2}
   m_i=\mu_i\left(1+\frac{g_9}{4}\mu_i^2+\frac{g_{10}}{4}\mu^2\right)
   +\frac{\kappa_1}{2}t_{ijk}\mu_j\mu_k.
\end{equation}
For the set $g_9=g_{10}=\kappa_1=0$ the current quark mass $m_i$ coincides
precisely  with the explicit symmetry breaking parameter $\mu_i$.

Note that the factor multiplying $h_i$ in the third term of eq. (\ref{h}) is
the same for each flavor. This quantity also appears in all meson mass
expressions, and there is no further dependence on the couplings $G, g_1, g_4$
involved for meson states with $a=1,2,\ldots,7$. Thus there is a freedom of
choice which allows to vary these couplings, condensates and quark masses
$\mu_i$, without altering this part of the meson mass spectrum.

To obtain the coefficients $h_{ab}^{(i)}$, $(i=1,2)$ in the Lagrangian $L_{aux}$
(\ref{lam}), it is sufficient to collect in the stationary phase equations
(\ref{sp}) only the terms linear in the fields, as can be seen from the
structure of the solutions (\ref{st}). Moreover, for any coefficient
multiplying a certain number $n$ of fields in $L_{aux}$ it is required to
consider terms only up to order $n-1$ in fields in the expansion (\ref{st}).
For instance, the inverse matrices to $h_{ab}^{(1)}$ and $h_{ab}^{(2)}$ are
\begin{eqnarray}
\label{h1}
   &&-2\left(h_{ab}^{(1)}\right)^{-1}= \left(2G+g_1h^2+g_4\mu h\right)\delta_{ab}
   +4g_1h_ah_b \nonumber \\
   &&+3A_{abc}\left(\kappa h_c+2\kappa_2\mu_c\right)
   +g_2h_{r}h_{c}\left(d_{abe}d_{cre}+2d_{ace}d_{bre}\right)
   \nonumber \\
   &&+g_3\mu_{r}h_{c}\left(d_{abe}d_{cre}+d_{ace}d_{bre}+d_{are}d_{bce}
   \right) \nonumber \\
   &&+2g_4\left(\mu_ah_b+\mu_bh_a\right)
   +g_5\mu_r\mu_c\left(d_{are}d_{bce}-f_{are}f_{bce}\right)
   \nonumber \\
   &&+g_6\mu_r\mu_cd_{abe}d_{cre}+4g_7\mu_a\mu_b.
\end{eqnarray}
\begin{eqnarray}
\label{h2}
   &&-2\left(h_{ab}^{(2)}\right)^{-1}= \left(2G+g_1h^2+g_4\mu h\right)\delta_{ab}
   \nonumber \\
   &&-3A_{abc}\left(\kappa h_c+2\kappa_2\mu_c\right)
   +g_2h_{r}h_{c}\left(d_{abe}d_{cre}+2f_{are}f_{bce}\right)
   \nonumber \\
   &&+g_3\mu_{r}h_{c}\left(d_{abe}d_{cre}+f_{are}f_{bce}+f_{ace}f_{bre}
   \right) \nonumber \\
   &&-g_5\mu_r\mu_c\left(d_{are}d_{bce}-f_{are}f_{bce}\right)
   \nonumber \\
   &&+g_6\mu_r\mu_cd_{abe}d_{cre}-4g_8\mu_a\mu_b.
\end{eqnarray}
These coefficients are totally defined in terms of $h_a$ and the parameters of
the model. Eqs. (\ref{h1})-(\ref{h2}) can be easily converted into explicit
formulae for $h_{ab}^{(i)}$, $(i=1,2)$.

Finally, to obtain the $h_{abc}^{(i)}$, $(i=1,2,3)$, of the interactions
involving three fields in $L_{aux}$, one equates the factors of $\sigma_a
\sigma_b$, $\phi_a\phi_b$, $\phi_a\sigma_b$ in (\ref{sp}) independently to zero.
After some algebra, this results into the following expressions
\begin{eqnarray}
\label{habc1}
    h_{abc}^{(1)}
    &=&\left[\frac{3\kappa}{4} A_{\bar a \bar b \bar c} + g_1(h_{\bar a}
       \delta_{\bar b \bar c} +2 h_{\bar c} \delta_{\bar a \bar b})
       \right. \nonumber \\
    &+&g_2 h_{\bar r} (d_{\bar a \bar b \bar \rho}d_{\bar r \bar c \bar \rho}
       +\frac{1}{2}d_{\bar a \bar r \bar \rho}d_{\bar b \bar c \bar \rho})
       \nonumber \\
    &+&\frac{g_3}{4}m_{\bar r}(2d_{\bar a\bar c\bar\rho}d_{\bar b \bar r \bar\rho}
       +d_{\bar b \bar c \bar\rho}d_{\bar a \bar r \bar \rho}-f_{\bar b \bar c \bar\rho}
       f_{\bar a \bar r \bar \rho})
       \nonumber \\
    &+&\left.\frac{g_4}{2}(m_{\bar a}\delta_{\bar b\bar c} +2m_{\bar c}
       \delta_{\bar a\bar b})\right]
       h^{(1)}_{a\bar a} h^{(1)}_{b\bar b} h^{(1)}_{c\bar c}
\end{eqnarray}
\begin{eqnarray}
\label{habc2}
   h_{abc}^{(2)}
   &=&\left[-\frac{3\kappa}{4} A_{\bar a\bar b\bar c}
      +g_1h_{\bar a}\delta_{\bar b \bar c}
      \right.\nonumber \\
   &+&g_2h_{\bar r}(f_{\bar a\bar b\bar\rho}f_{\bar c\bar r\bar\rho}
      +\frac{1}{2}d_{\bar a\bar r\bar\rho}d_{\bar b\bar c\bar\rho})
      \nonumber \\
   &-&\frac{g_3}{4}m_{\bar r}(2f_{\bar a\bar c\bar\rho}f_{\bar b\bar r\bar\rho}
      +f_{\bar b\bar c\bar\rho}f_{\bar a\bar r\bar\rho}-d_{\bar b\bar c\bar\rho}
      d_{\bar a\bar r\bar\rho})\nonumber \\
   &+&\left.\frac{g_4}{2}m_{\bar a}\delta_{\bar b\bar c}\right]
      h^{(1)}_{a\bar a}h^{(2)}_{b\bar b}h^{(2)}_{c\bar c}
\end{eqnarray}
\begin{eqnarray}
\label{habc3}
   h_{abc}^{(3)}
   &=&\left[-\frac{3\kappa}{2}A_{\bar a\bar b\bar c}
      +2g_1h_{\bar c}\delta_{\bar b\bar a}
      \right.\nonumber \\
   &+&g_2h_{\bar r}(d_{\bar a\bar b\bar\rho}d_{\bar c\bar r\bar\rho}
      +f_{\bar r\bar a\bar\rho}f_{\bar c\bar b\bar\rho}
      +f_{\bar r\bar b\bar\rho}f_{\bar c\bar a\bar\rho})
      \nonumber \\
   &+&\frac{g_3}{2}m_{\bar r}(d_{\bar a\bar b\bar\rho}d_{\bar c\bar r\bar\rho}
      +f_{\bar b\bar c\bar\rho}f_{\bar a\bar r\bar\rho}+f_{\bar a\bar c\bar\rho}
      f_{\bar b\bar r\bar\rho})
      \nonumber \\
   &+&\left. g_4m_{\bar c}\delta_{\bar b\bar a}\right]
      h^{(2)}_{a\bar a}h^{(2)}_{b\bar b}h^{(1)}_{c\bar c}.
\end{eqnarray}
Contracting with $\phi_b\phi_c$ in eq. (\ref{lam}), one sees that the term
going with $h_{abc}^{(2)}$ is simply half the one going with $h_{bca}^{(3)}$, and
$L_{aux}$ simplifies to
\begin{eqnarray}
\label{lams}
   L_{aux}\!\!\!\!\!\!
   &&=h_a\sigma_a+\frac{1}{2}\,h_{ab}^{(1)}\sigma_a\sigma_b
      +\frac{1}{2}\,h_{ab}^{(2)}\phi_a\phi_b  \nonumber \\
   &&+\,\sigma_a\left(\,\frac{1}{3}\, h^{(1)}_{abc}\sigma_b\sigma_c
   +h^{(2)}_{abc}\phi_b\phi_c\right)
   + \ldots
\end{eqnarray}
Although there are five parameters $\kappa,g_1,g_2,g_3,g_4$ which appear
explicitly in $h_{abc}^{(i)}$, they do not represent new freedom to fit the
meson interaction dynamics, since they occur also in the $h_{ab}^{(i)}$;
through the latter the $h_{abc}^{(i)}$ depend implicitly also on further six
parameters $G,\kappa_2,g_5,g_6,g_7,g_8$. All will be fixed by fitting the mass
spectra and weak decay constants, see (\ref{mass}) and section IV below.

\subsection{The heat kernel contribution}

We now turn our attention to the total Lagrangian of the bosonized theory. To
write down this Lagrangian we should add the terms coming from integrating out
the quark degrees of freedom in (\ref{L}) to our result (\ref{lams}).
Fortunately, the technicalities are known. We use the modified heat kernel
technique \cite{Osipov:2001,Osipov:2001a,Osipov:2001b} developed for the case
of explicit chiral symmetry breaking. In the isospin limit one can find all
necessary details of such calculations for instance in \cite{Osipov:2006a}. For
future reference we apply it here to obtain the result for the more general
case in which the strong isospin symmetry is broken.

From the vacuum to vacuum persistence amplitude in the spontaneous broken phase
\begin{eqnarray}
\label{Z}
   Z[ \sigma,\phi ]
   &\!=\!&\int\! {\cal D}q{\cal D}\,\bar q\exp\left(i\!\int\! d^4x\,
      {\cal L}_q(\sigma,\phi )\right),
      \nonumber \\
   {\cal L}_q (\sigma,\phi )
   &\!=\!&\bar q\left(i\gamma^\mu\partial_\mu -M-\sigma -i\gamma_5\phi\right)q
\end{eqnarray}
the heat kernel result for the integration over the quark degrees of freedom is
\begin{eqnarray}
\label{W}
   W[Y]
   &\!=\!&\ln |\det D|=-\frac{1}{2} \int_0^\infty \frac{dt}{t} \rho(t)
   \exp\left(-t D_E^\dagger D_E\right),
   \nonumber \\
   D_E^\dagger D_E
   &\!=\!&M^2 -\partial^2 +Y, \quad Y=i\gamma_\mu (\partial_\mu +i\gamma_5
   \partial_\mu \phi )
   \nonumber \\
   &\!+\!&\sigma^2+\{M,\sigma\}+\phi^2+i\gamma_5[\sigma+M,\phi ],
\end{eqnarray}
or
\begin{equation}
\label{W1}
   W[Y]=-\int \frac{d^4 x_E}{32\pi^2} \sum_{i=0}^\infty I_{i-1} \mbox{tr}[b_i]
\end{equation}
where $D_E$ stands for the Dirac operator in Euclidean space. We consider the
expansion up to the third Seeley-DeWitt coefficient $b_i$
\begin{eqnarray}
\label{bi}
   b_0&\!=\!&1, \quad b_1=-Y,
   \nonumber \\
   b_2&\!=\!&\frac{Y^2}{2}+\frac{\lambda_3}{2} \Delta_{ud} Y +
   \frac{\lambda_8}{2 \sqrt{3}}(\Delta_{us}+\Delta_{ds}) Y,
\end{eqnarray}
with $\Delta_{ij}=M_i^2-M_j^2$. This order of the expansion takes into account
the dominant contributions of the quark one-loop integrals $I_i$ $(i=0,1,
\ldots )$; these are the arithmetic average values $I_i=\frac{1}{3}
[J_i(M_u^2)+J_i(M_d^2)+J_i(M_s^2)]$ where
\begin{equation}
\label{ji}
   J_i(m^2)=\int\limits_0^\infty\frac{{\rm d}t}{t^{2-i}}\rho
   (t\Lambda^2) e^{-t m^2},
\end{equation}
with the Pauli-Villars regularization kernel \cite{Osipov:2004a,Osipov:2004b}
\begin{equation}
\label{PVk}
   \rho (t\Lambda^2)=1-(1+t\Lambda^2)\exp (-t \Lambda^2).
\end{equation}
In the following we need therefore only to know two of them (the lowest order
$\sim b_0$ contributes to the effective potential and is not needed in the
present study)
\begin{equation}
\label{j0}
   J_0(m^2)=\Lambda^2- m^2\ln\left(1+\frac{\Lambda^2}{m^2}\right),
\end{equation}
and
\begin{equation}
\label{j1}
      J_1(m^2)=\ln\left(1+\frac{\Lambda^2}{m^2}\right)
      -\frac{\Lambda^2}{\Lambda^2+m^2}\ .
\end{equation}

While both terms proportional to $b_1$ and $b_2$ have contributions to the gap
equations and meson masses, only $b_2$ contributes to the kinetic and
interaction terms. The $\sigma$ tadpole term must be excluded from the total
Lagrangian. This gives us a system of gap equations
\begin{equation}
\label{gap}
   h_i+\displaystyle\frac{N_c}{6\pi^2} M_i
   \left[3I_0-\left(3M_i^2-M^2\right) I_1 \right]=0.
\end{equation}
Here $N_c=3$ is the number of colors, and $M^2=M_u^2+M_d^2+M_s^2$. Combining all
terms of the total Lagrangian $L=L_{kin}+ L_{mass}+L_{int}$ that contribute to
the kinetic terms $L_{kin}$ and meson masses $L_{mass}$ one gets
\begin{eqnarray}
\label{mass}
     &&L_{kin}+L_{mass}  \nonumber \\
     &\!=\!&\frac{N_cI_1}{16\pi^2}\,\mbox{tr}\left[
           (\partial_\mu \sigma )^2+(\partial_\mu \phi )^2\right]
           +\frac{N_cI_0}{4\pi^2}(\sigma_a^2+\phi_a^2)
           \nonumber \\
     &\!-\!&\frac{N_cI_1}{12\pi^2}
           \left\{
           \left[2\left(M_u+M_d\right)^2-M_u M_d -M_s^2\right](\sigma_1^2
           +\sigma_2^2)
           \right.
           \nonumber \\
     &\!+\!&\left[2\left(M_u+M_s\right)^2-M_u M_s -M_d^2\right]
           \left(\sigma_4^2+\sigma_5^2\right)
           \nonumber \\
     &\!+\!&\left[2(M_d+M_s)^2-M_dM_s-M_u^2\right]
           \left(\sigma_6^2+\sigma_7^2\right)
           \nonumber \\
     &\!+\!&\frac{1}{2}\left[\sigma_u^2\left(8 M_u^2-M_d^2 -M_s^2\right)
           +\sigma_d^2\left(8 M_d^2-M_u^2 -M_s^2\right)
           \right.
           \nonumber \\
     &\!+\!&\left.\sigma_s^2\left(8 M_s^2-M_u^2 -M_d^2\right)\right]
           \nonumber \\
     &\!+\!&\frac{1}{2}\left[\phi_u^2\left(2 M_u^2-M_d^2 -M_s^2\right)
           +\phi_d^2\left(2 M_d^2-M_u^2 -M_s^2\right)
           \right.
           \nonumber \\
     &\!+\!&\left.\phi_s^2\left(2 M_s^2-M_u^2 -M_d^2\right)\right]
           \nonumber \\
     &\!+\!&\left[2 \left(M_u-M_d\right)^2+M_uM_d-M_s^2\right]
           \left(\phi_1^2+\phi_2^2\right)
           \nonumber \\
     &\!+\!&\left[2 \left(M_u-M_s\right)^2+M_u M_s -M_d^2\right]
           \left(\phi_4^2+\phi_5^2\right)
           \nonumber \\
     &\!+\!&\left.\left[2\left(M_d-M_s\right)^2+M_dM_s-M_u^2\right]
           \left(\phi_6^2+\phi_7^2\right)\right\}
           \nonumber \\
     &\!+\!&\frac{1}{2}\,h_{ab}^{(1)}\sigma_a\sigma_b
     +\frac{1}{2}\,h_{ab}^{(2)}\phi_a\phi_b.
\end{eqnarray}
The kinetic term requires a redefinition of meson fields,
\begin{equation}
\label{ren}
   \sigma_a =g\sigma_a^R, \quad \phi_a =g\phi_a^R, \quad
   g^2=\frac{4\pi^2}{N_cI_1},
\end{equation}
to obtain the standard factor $1/4$. The flavor and charged fields are related
through
\begin{eqnarray}
   &&\frac{\lambda_a}{\sqrt{2}}\phi_a =
     \pmatrix{
             \frac{\phi_u}{\sqrt{2}} &\pi^+& K^+ \cr
             \pi^- &\frac{\phi_d}{\sqrt{2}}& K^0 \cr
             K^- &{\bar K}^0 &\frac{\phi_s}{\sqrt{2}} \cr
             }
   \nonumber\\
   &&\frac{\lambda_a}{\sqrt{2}}\sigma_a =
     \pmatrix{
             \frac{\sigma_u}{\sqrt{2}} &a_0^+& \kappa^+ \cr
             a_0^-&\frac{\sigma_d}{\sqrt{2}}& \kappa^0  \cr
             \kappa^-&{\bar \kappa}^0&\frac{\sigma_s}{\sqrt{2}}\cr
             }
\end{eqnarray}
and in particular for the diagonal components
\begin{eqnarray}
\label{dico}
     \phi_u&=&\phi_3+\frac{\sqrt{2}\phi_0 +\phi_8}{\sqrt{3}}
            =\phi_3+\eta_{ns} \nonumber \\
     \phi_d&=&-\phi_3+\frac{\sqrt{2} \phi_0 +\phi_8}{\sqrt{3}}
            =-\phi_3+\eta_{ns} \nonumber \\
     \phi_s&=&\sqrt{\frac{2}{3}}\phi_0-\frac{2 \phi_8}{\sqrt{3}}
            =\sqrt{2}\eta_s
\end{eqnarray}
and similar for the scalar fields. Here we also introduce the $\eta_{ns}$ and
$\eta_s$ which stand for the flavor components of the physical $\eta, \eta'$
states in the nonstrange and strange basis. In addition to the flavor mixing in
the $\eta, \eta'$ channels the isospin breaking induces a coupling between the
$\pi^0$ and these states
\begin{equation}
\label{pee}
     \pi^0=\phi_3+\epsilon\eta +\epsilon'\eta'.
\end{equation}
To get the physical $\pi^0$, $\eta$ and $\eta'$ mesons and correspondingly the
scalar $a_0^0(980)$, $\sigma$ and $f_0(980)$ mesons one may proceed as in
\cite{Kroll:2005}. Since $\phi_3$ couples weakly to the $\eta_{ns}$ and $\eta_s$
states (decoupling in the isospin limit) while the $\eta-\eta'$ mixing is
strong, it is appropriate to use isoscalar $\eta_{ns}, \eta_s$ and isovector
$\phi_3$ combinations as a starting point for an unitary transformation to the
physical meson states $\pi^0, \eta, \eta'$. In this case the corresponding
unitary matrix ${\cal U}$ can be linearized in the $\pi^0-\eta$ and $\pi^0-
\eta'$ mixing angles $\epsilon_1, \epsilon_2\sim {\cal O}(\epsilon ), \epsilon
\ll 1$. Precisely \cite{Kroll:2005}
\begin{equation}
    \left( \begin{array}{c} \pi^0 \\
                            \eta  \\
                            \eta'
           \end{array} \right)
    = {\cal U}(\epsilon_1,\epsilon_2,\psi )
\left( \begin{array}{c} \phi_3 \\
                        \eta_{ns}  \\
                        \eta_{s}
           \end{array} \right),
\end{equation}
where
\begin{equation}
   {\cal U} = \left(\begin{array}{ccc} 1& \epsilon_1 +\epsilon_2\cos\psi
                                & -\epsilon_2\sin\psi \\
            -\epsilon_2-\epsilon_1\cos\psi& \cos\psi& -\sin\psi \\
            -\epsilon_1\sin\psi & \sin\psi& \cos\psi
            \end{array} \right)
\end{equation}
In particular, in eq.(\ref{pee}) $\epsilon =\epsilon_2 +\epsilon_1\cos\psi,
\epsilon'=\epsilon_1\sin\psi$.

In the isospin limit we use the mixing angle conventions summarized in the
Appendix B of \cite{Osipov:2004b}. We have the following different
possibilities of relating the physical states $(\bar{X},X)$ with the states of
the strange-nonstrange basis
\begin{equation}
\label{basisnss}
   \left(\begin{array}{c} \bar X \\ X
         \end{array} \right)
   =R_\psi
   \left(\begin{array}{c} X_{ns} \\ X_s
         \end{array} \right)
   =R_{\bar\psi}
   \left(\begin{array}{c} -X_s   \\ X_{ns}
         \end{array} \right),
\end{equation}
where the orthogonal $2\times 2$ matrix $R_\psi$ is
\begin{equation}
   R_\psi =
   \left(\begin{array}{cc} \cos\psi & -\sin\psi \\
                           \sin\psi & \cos\psi
   \end{array} \right),
\end{equation}
or of the singlet-octet basis
\begin{equation}
\label{basis08}
   \left(\begin{array}{c} \bar X \\ X
         \end{array} \right)
   =R_\theta
   \left(\begin{array}{c} X_8 \\ X_0
         \end{array} \right).
\end{equation}
Here $\theta$, being a solution of the equation $\tan 2\theta =x$, is the
principal value of $\mbox{arctan}\,x$, i.e. belongs to the interval
$-(\pi /4)\le\theta\le (\pi /4)$. The angle $\psi$ is related with $\theta$ by
the equation $\psi=\theta + {\bar{\theta}}_{id}$, where $\bar\theta_{id}$ ($
\theta_{id}+\bar{\theta}_{id}=\pi /2$) is determined by the equations $\sin\bar
\theta_{id}=\sqrt{2/3}$, $\cos\bar\theta_{id}=1/\sqrt{3}$, therefore $\psi =
\theta +\mbox{arctan}\sqrt{2}=\theta +54.74^\circ$. It means that $\psi$ is
restricted to the range $9.74^\circ\le\psi\le 99.74^\circ$. If the value of
$\psi$ leaves the range, we must resort to the angle $\bar\psi =\psi - (\pi /2)
=\theta -\theta_{id}$, taking values in the interval $-80.26^\circ\le\bar\psi\le
9.74^\circ$. These two angles correspond to two alternative phase conventions
for a strange $\bar ss$-component. As a result of the following numerical
calculations, in the case of the pseudoscalars the identification of the
physical states is $\bar X=\eta,\, X=\eta'$ and for the scalars $\bar
X=f_0(980),\, X=\sigma$.

We turn to the interaction terms of the heat kernel action in (\ref{W}). The
only contribution comes from $Y^2/2$ in the term proportional to $b_2$ and
reads
\begin{eqnarray}
\label{hkint}
   L^{(hk)}_{int}&=& -\frac{N_c}{2 \pi^2} I_1 M_a
   \left[d_{ab\rho}d_{ce\rho}\sigma_b\left(\sigma_c\sigma_e+\phi_c\phi_e\right)
   \right. \nonumber \\
   &+&\left. 2f_{ac\rho}f_{be\rho}\sigma_b\phi_c\phi_e\right],
\end{eqnarray}
which must be added to the interaction piece stemming from (\ref{lams}),
yielding the total interaction Lagrangian
\begin{equation}
\label{lint}
   L_{int}=L^{(hk)}_{int} +\sigma_a\left(\frac{1}{3} h^{(1)}_{abc}\sigma_b\sigma_c
   +h^{(2)}_{abc}\phi_b\phi_c\right).
\end{equation}
Note that all dependence on the parameters of the explicit symmetry breaking
quark interactions is explicitly absorbed in the bosonized Lagrangian through
the matrices $h_{ab}^{(1,2)}$ for the meson mass spectra (\ref{mass}) and through
the $h_{abc}^{(1,2,3)}$ for the meson interaction Lagrangian (\ref{lint}). In
other words, the formal structure of the Lagrangian (\ref{lams}) in comparison
to the case without these interactions remains unchanged. This differs from the
heat kernel Lagrangian where the information about the difference in
constituent quark masses leads to a resummation of the heat kernel series for
the modified Seeley-DeWitt coefficients $b_i$ \cite{Osipov:2001,Osipov:2001a,Osipov:2001b}.
The parameters of these two seemingly
separated sectors of the Lagrangian, i.e. the constituent quark masses and
scale parameter $\Lambda$ for the heat kernel Lagrangian on one hand, and the
multiquark interaction couplings for the SPA piece on the other hand, are
connected through the gap equations (\ref{gap}) which must be solved
self-consistently with the SPA equations (\ref{h}).

In the remaining of this subsection we discuss the scheme in which the strong
decay widths of the scalar mesons are calculated. Given the complexity of the
Lagrangian, we will restrict our study of the decays to the tree level bosonic
couplings (\ref{hkint}), (\ref{lint}). To deal in an approximate way with the
proximity of particle thresholds to the resonance mass we shall resort to the
widely accepted Flatt\'e type distribution \cite{Flatte:1976}. Other closed
bosonic channel contributions will not be taken into consideration for
simplicity, since the ratios of couplings in the concurring closed channels to
the nominal one turn out to be numerically less relevant in our fits.

The strong decay width of the scalar meson $S$ in two pseudoscalars $P_1,P_2$
are thus obtained as
\begin{equation}
\label{strdec}
   \Gamma_\beta =\frac{|\vec{p_\beta}|}{8\pi m_S^2}|g_\beta|^2
               \equiv {\bar g}_{\beta}|\vec{p}_\beta|
\end{equation}
with
$$
   |{\vec p}_\beta|=\sqrt{\frac{\left[m_S^2-(m_1+m_2)^2\right]
                   \left[m_S^2-(m_1-m_2)^2\right]}{4m_S^2}}
$$
where index $\beta$ specifies all necessary kinematic characteristics of the
channel $S\to P_1P_2$, and the masses $m_S, m_1, m_2$ of the states. We
introduce also a shorthand notation for the dimensionless quantity
${\bar g}_{\beta}$ in eq.(\ref{strdec}). In this definition we include all
flavor and symmetry factors associated with the final state.

The so obtained widths are valid in the Breit-Wigner resonance scheme, which is
known to be an incomplete description for decays with the resonance mass close
to the threshold of particle emission. We use Flatt\'e distributions in the
cases of the $a_0(980)$ and $f_0(980)$ decays to accomodate the threshold
effects associated with the two kaon production, on grounds of analyticity and
unitarity at the threshold. Close to this threshold the elastic scattering
cross section for $\pi\eta$ in the case of $a_0$ or $\pi\pi$ for $f_0$ is
parametrized by a two-channel resonance
\begin{eqnarray}
\label{flatte}
   \sigma_{el}&\!=\!&4\pi|f_{el}|^2,
   \nonumber \\
   f^{\beta}_{el}&\!=\!&\frac{1}{|\vec{p}_\beta|}\frac{m_R \Gamma_{\beta}}{m_R^2
   -s-im_R(\Gamma_{\beta} +\Gamma^S_{K\bar K})}
\end{eqnarray}
with the index $\beta$ designating here either the $a_0\pi\eta$ or the
$f_0\pi\pi$ channels
and
\begin{equation}
   \Gamma^S_{K\bar K}=\left\{
   \begin{array}{rcl}
   {\bar g}^S_K\sqrt{\frac{s}{4}-m_K^2} &\hspace{0.5cm}&
   \mbox{above threshold} \\
   {i\bar g}^S_K\sqrt{m_K^2-\frac{s}{4}} &\hspace{0.5cm}&
   \mbox{below threshold}.\\
   \end{array}\right.
\end{equation}
where  ${\bar g}^S_K$ stands for the coupling of $S$ to the two kaons, in this
case $S=a_0$ or $f_0$. Here $m_R$ is the nominal resonance mass and
$s=(p_1+p_2)^2$, where $p_1,p_2$ are the 4-momenta of $P_1$ and $P_2$. Near the
${K\bar K}$ threshold only the width $\Gamma^S_{K\bar K}$ is expected to vary
strongly; the widths $\Gamma_\beta$ are approximated by a constant value in
this region, taken to be (\ref{strdec}) evaluated at $s=m_R^2$, since the
$\pi\eta$ and $\pi\pi$ thresholds lie further away from the resonance. The
numerical results are presented and discussed in the section IV.

\subsection{A note on radiative decays}

Additional information on the structure of the mesons is obtained through the
study of their radiative decays. We consider in this work the two photon decays
at the quark one-loop order of the scalar and pseudoscalar mesons. The
corresponding integrals are finite. A direct extension of the heat kernel
Lagrangian to incorporate the coupling to the electromagnetic interaction shows
that there is no contribution up to the order $b_2$ of the Seeley-DeWitt
coefficients for the scalar decays. The anomalous pseudoscalar - two photon
decays belong to the imaginary part of the action and are not contemplated by
the heat kernel techniques considered, which apply only to the real part. By
the Adler-Bardeen theorem \cite{Adler:1969,Bell:1969,AdlerBardeen:1969} they
are fully determined by the three-point function Feynman amplitudes involving
one quark loop; higher orders only redefine the couplings. There is however a
source of uncertainty which resides in the model dependent determination of the
coupling of the $\eta$ and $\eta'$ mesons to the quarks. In our approach they
are calculated within the heat kernel technique outlined in section III.B.
Regarding the scalar meson two photon decays, they are also most simply
evaluated through the three-point Feynman amplitudes, keeping only the
contribution corresponding to the first non-vanishing order in the heat kernel
action, that is the term involving the Seeley-DeWitt coefficient $b_3$. From
now on we will consider the case with exact $SU(2)$ isospin symmetry, i.e.
$\mu_u=\mu_d=\hat\mu\neq \mu_s$, and $M_u=M_d=\hat M\neq M_s$. With the standard
electromagnetic coupling to quarks ${\cal L}_\gamma=-e{\bar q}\gamma^\mu Q q
A_\mu$, $Q=\frac{1}{2}(\lambda_3+\frac{1}{\sqrt{3}} \lambda_8)$ and using the
Pauli-Villars regularization, the scalar meson photon photon amplitude $A$:
$S(s)\rightarrow\gamma(p_1,\epsilon^*_\mu)+\gamma(p_2,\epsilon^*_\nu)$  is
obtained in terms of the gauge invariant tensor ${\cal L}_{\mu\nu}= (p_2^\mu
p_1^\nu-\frac{1}{2} s g^{\mu\nu})$, with $s=(p_1+p_2)^2$
\begin{eqnarray}
\label{SVV}
   A_{S\gamma\gamma}^{\mu\nu}
   &=&{\cal L}^{\mu\nu} A_{S\gamma\gamma}; \qquad
      S=\sigma,f_0(980),a_0(980) \nonumber \\
   A_{\sigma\gamma\gamma}
   &=&\frac{5}{9}\,T_u\cos\bar{\psi}-\frac{\sqrt{2}}{9}\,T_s \sin\bar{\psi}
   \nonumber \\
   A_{f_0\gamma\gamma}
   &=&-\frac{5}{9}\,T_u\sin\bar{\psi}-\frac{\sqrt{2}}{9}\,T_s\cos\bar{\psi}
   \nonumber \\
   A_{a_0\gamma\gamma}
   &=&\frac{1}{3}\,T_u
\end{eqnarray}
where
\begin{eqnarray}
   T_i
   &=&32\pi\alpha g M_i Q_3(s,M_i), \quad \mbox{i=(u,s)} \nonumber \\
   Q_3(s,M_i)
   &=&\frac{iN_c}{16\pi^2}\int_0^1 dx \int_0^{1-x}dy (1-4xy) \nonumber \\
   &\times& \int_0^\infty dt \rho(t\Lambda^2) e^{-t(M_i^2-xys)}
\end{eqnarray}
$\alpha=\frac{e^2}{4\pi}$ is the fine structure constant and $g$ the field
normalization defined in (\ref{ren}). The factors of $T_i$ result from the
flavor traces and projection to the physical states with the angle $\bar\psi$
defined in (\ref{basisnss}). The result for the integral $Q_3(s,M_i)$ with the
Pauli-Villars kernel $\rho(t\Lambda^2)$, eq. (\ref{PVk}), has been evaluated in
\cite{Osipov:1996}. To obtain the dominant contribution, i.e. the first
non-vanishing order in the heat kernel series, one needs to express the
integrals $Q_3(s,M_i)$ as the following averaged sum evaluated at $s=0$
\cite{Osipov:2001a,Osipov:2001b}
\begin{eqnarray}
   Q_3(0,M_i)&\rightarrow & Q_3(0,M_u,M_s) \nonumber \\
   &=&\frac{1}{3}(2 Q_3(0,M_u)+Q_3(0,M_s)) \nonumber \\
   &+&{\cal O}(b_3)
\end{eqnarray}
where the term ${\cal O}(b_3)$ is discarded as it belongs to the next order in
the heat kernel series (\ref{W}), and
\begin{equation}
\label{Q30}
   Q_3(0,M_i)=-\frac{N_c}{48\pi^2 M_i^2}\left(\frac{\Lambda^2}{\Lambda^2+M_i^2}
   \right)^2,
\end{equation}
or, in the notation of (\ref{ji}), we have that
\begin{equation}
   Q_3(0,M_i)=-\frac{N_c}{48\pi^2}J_2(M_i^2).
\end{equation}
Finally the decay widths for the scalar mesons in the narrow width
approximation are given as (see also \ref{dpgg})
\begin{equation}
\label{dsgg}
   \Gamma_{S\gamma\gamma}=\frac{m_S^3}{64 \pi} |A_{S\gamma\gamma}|^2
\end{equation}

The anomalous decay of the pseudoscalars $P=(\pi^0,\eta,\eta')$ in two photons
$P(p)\rightarrow\gamma(p_1,\epsilon^*_\mu)+\gamma(p_2,\epsilon^*_\nu)$ has the
same Lorentz structure in all channels and reads
\begin{eqnarray}
\label{PVV}
   A_{P\gamma\gamma}^{\mu\nu}
   &=&\epsilon^{\mu\nu\alpha\beta}p_{1\alpha}p_{2\beta} A_{P\gamma\gamma}
   \nonumber \\
   A_{\eta\gamma\gamma}
   &=&-\frac{5}{9}\,T^P_u\sin\bar{\psi}_P-\frac{\sqrt{2}}{9}\,
   T^P_s\cos\bar{\psi}_P \nonumber \\
   A_{\eta'\gamma\gamma}
   &=&\frac{5}{9}\,T^P_u\cos\bar{\psi}_P-\frac{\sqrt{2}}{9}\,T^P_s
   \sin\bar{\psi}_P \nonumber \\
   A_{\pi^0\gamma\gamma}
   &=&\frac{1}{3}\,T^P_u
\end{eqnarray}
where $\bar{\psi}_P$ stands for the mixing angle in the pseudoscalar channels,
eq. (\ref{basisnss}) and
\begin{eqnarray}
\label{TP}
   T^P_i(s,M_i)
   &=&32\pi\alpha g M_i I_P(s,M_i) \nonumber \\
   I_P(s,M)
   &=&\frac{-N_c}{16\pi^2}\int_0^1 dx \int_0^{1-x}dy \int_0^\infty dt
   e^{-t(M^2-xys)} \nonumber \\
\end{eqnarray}
and the contribution to the imaginary part of the heat kernel action is
\begin{equation}
I_P(0,M)=\frac{-N_c}{32\pi^2 M^2}.
\end{equation}
At this stage one sees that the only parameter dependence in the radiative
decays of the scalars and pseudoscalars enters through the wave function
normalization $g$, common to all decays considered, and through the constituent
quark masses; there is also an explicit dependence on the scale $\Lambda$ in
the case of the scalar decays through the factor
$(\frac{\Lambda^2}{\Lambda^2+M^2})^2$ in (\ref{Q30}). The PCAC hypothesis
establishes a relation between $g$, the weak pion and kaon decay couplings and
the constituent quark masses (see also (\ref{rm}) below)
\begin{equation}
\label{wd}
   f_\pi=\frac{\hat M}{g}; \qquad f_K=\frac{{\hat M} +M_s}{2g}.
\end{equation}
These identities allow to eliminate all dependence on the constituent quark
masses from the pseudoscalar radiative decays, leading to
\begin{equation}
\label{tp}
   T^P(0,\hat M)=\frac{N_c \alpha}{\pi f_\pi}, \quad
   T^P(0,M_s)=\frac{N_c \alpha}{\pi (2f_K-f_\pi)}.
\end{equation}
One obtains then the celebrated relation $A_{\pi\gamma\gamma}=\frac{\alpha}{\pi
f_\pi}$ for the $\pi^0$ decay amplitude \cite{Adler:1969}. The Adler-Bardeen
theorem allows to infer that the study and measurement of the anomalous decays
are a reliable means of determination of the  mixing angle of the $\eta$ and
$\eta'$ mesons, which must comply with the mixing angle determination extracted
from the mass spectrum. One should also stress that with the present model
Lagrangian one is able to account properly for the $SU(3)$ breaking effects in
the description of the weak decay constants $f_\pi$ and $f_K$, in addition to
having the correct empirical $\eta$ and $\eta'$ meson masses (see section IV),
which has been an open problem until now. This is important for the numerical
consistency in the amplitudes (\ref{tp}).

The respective widths are calculated as
\begin{equation}
\label{dpgg}
   \Gamma_{P\gamma\gamma}=\frac{|\vec{p}|^3}{8 \pi} |A_{P\gamma\gamma}|^2
\end{equation}
with $|\vec{p}|=\sqrt{m_P^2/4}$ and $m_P$ the pseudoscalar mass. The numerical
results are presented in section IV.

\section{Fixing parameters, numerical results and discussion}

\subsection{Meson Spectra and weak decays}

In the chiral limit, $m_u=m_d=m_s=0$, the Lagrangian (\ref{mass}) leads to the
conserved vector, ${\cal V}_\mu^a$, and axial-vector, ${\cal A}_\mu^a$,
currents. The matrix elements of axial-vector currents
\begin{equation}
      \langle 0|{\cal A}_\mu^a (0)|\phi^b_R(p)\rangle = ip_\mu f^{ab}
\end{equation}
define the weak and electromagnetic decay constants of physical pseudoscalar
states (see details in \cite{Osipov:2006a}). Now let us fix the values of the
various quantities introduced. After choosing the set $\kappa_1=g_9=g_{10}=0$
we still have to fix fourteen parameters: $\Lambda, \hat m, m_s, G, \kappa,
\kappa_2$ and $g_1,\ldots,g_8$. There are two intrinsic restrictions of the
model, namely, the stationary phase (\ref{h}) and the gap (\ref{gap})
equations, which as mentioned above must be solved self-consistently. This is
how the explicit symmetry breaking is intertwined with the dynamical symmetry
breaking and vice versa. We use (\ref{gap}) to determine $\hat h, h_s$ through
$\Lambda, M_s$ and $\hat M$. The ratio $M_s/\hat M$ is related to the ratio of
the weak decay constants of the pion, $f_\pi=92$\ MeV, and the kaon,
$f_K=113$\ MeV. Here we obtain
\begin{equation}
\label{rm}
   \frac{M_s}{\hat M}=2\frac{f_K}{f_\pi}-1=1.46.
\end{equation}

Furthermore, the two eqs. (\ref{h}) can be used to find the values of
$\Lambda$ and $\hat M$ if the parameters $\hat m$, $m_s$, $G$, $\kappa$,
$\kappa_2$, $g_1, \ldots, g_7$ are known. Thus, together with $g_8$ we have at
this stage thirteen couplings to be fixed. Let us consider the current quark
masses $\hat m$ and $m_s$ to be an input. Their values are known, from various
analyses of the chiral treatment of the light pseudoscalars, to be around
$\hat m=4$\ MeV and $m_s=100$\ MeV \cite{PDT:2011}. Then the remaining eleven
couplings can be found by comparing with empirical data. One should stress the
possibility (which did not exist before the inclusion of mass-dependent
interactions) to fit the low lying pseudoscalar spectrum, $m_\pi=138$\ MeV,
$m_K=494$\ MeV, $m_\eta=547$\ MeV, $m_{\eta'}=958$\ MeV, the weak pion and kaon
decay constants, $f_\pi=92$\ MeV, $f_K=113$\ MeV, and the singlet-octet mixing
angle $\theta_p=-15^\circ$ to perfect accuracy, see Table \ref{table-1}.

One can deduce that the couplings $\kappa_2$ and $g_8$ are essential to improve
the description in the pseudoscalar sector; in particular, $g_8$ is responsible
for fine tuning the $\eta\!-\!\eta'$ mass splitting, see also Table II, where
the difference in $g_8$ between set (b) and sets (a,c,d) is due to the input
$\theta_P=-15^\circ$ versus $\theta_P=-12^\circ$ respectively.

The remaining five conditions are taken from the scalar sector of the model.
Unfortunately, the scalar channel in the region about $1$\ GeV became a
long-standing problem of QCD. The abundance of meson resonances with $0^{++}$
quantum numbers shows that one can expect the presence of non-$q\bar q$ scalar
objects, like glueballs, hybrids, multiquark states and so forth
\cite{Klempt:2007}. This creates known difficulties in the interpretation and
classification of scalars. For instance, the numerical attempts to organize the
$U(3)$ quark-antiquark nonet based on the light scalar mesons, $\sigma$ or
$f_0(600),$ $a_0(980),$ $\kappa (800), f_0(980)$, in the framework of NJL-type
models have failed (see, e.g. \cite{Weise:1990,Vogl:1990,Weise:1991,Volkov:1984,Volkov:1986,Osipov:2004b,Su:2007}). The reason is the ordering of the
calculated spectrum which typically is $m_\sigma<m_{a_0}<m_\kappa <m_{f_0}$, as
opposed to the empirical evidence: $m_\kappa<m_{a_0}\simeq m_{f_0}$.

On the other hand, it is known that a unitarized nonrelativistic meson model
can successfully describe the light scalar meson nonet as $\bar qq$ states
with a meson-meson admixture \cite{Beveren:1986}. Another model which assumes
the mixing of $q\bar q$-states with others, consisting of two quarks and two
antiquarks, $q^2\bar q^2$ \cite{Jaffe:1977}, yields a possible description of
the $0^{++}$ meson spectra as well \cite{Schechter:2008,Schechter:2009}. The
well known model of Close and T\"ornqvist \cite{Close:2002} is also designed
to describe two scalar nonets (above and below $1$\, GeV). The light scalar
nonet below $1$\, GeV has a core made of $q^2\bar q^2$ states with a small
admixture of a $\bar qq$ component, rearranged asymptotically as
meson-meson states. These successful solutions seemingly indicate on the
importance of certain admixtures for the correct description of the light
scalars. Our model contains such admixtures in the form of the appropriate
effective multi-quark vertices with the asymptotic meson states described by
the bosonized $\bar qq$ fields. We have found, that the quark mass
dependent interactions can solve the problem of the light scalar spectrum and
these masses can be understood in terms of spontaneous and explicit chiral
symmetry breaking only. Indeed, one can easily fit the data: $m_\sigma =600$\
MeV, $m_{a_0}=980$\ MeV, $m_\kappa =850$\ MeV, $m_{f_0}=980$\ MeV. In this case
we obtain for the singlet-octet mixing angle $\theta_s$ roughly $\theta_s=
19^\circ$ \cite{Osipov:2013}. Without changing the mass spectra better fits for
the strong radiative decays of the scalars are obtained with
$\theta_s=25^\circ\div 28^\circ$, in the next subsection.

\begin{table*}
\caption{The same values for the pseudoscalar and scalar masses (except for
$m_\sigma$) and weak decay constans (all in MeV) are used as input (marked with *) for
different sets of the model. Parameter sets (a),(b),(c),(d) of all following
tables differ by varying the mixing angles and $m_\sigma$: sets (a), (b) and (d)
with $m_\sigma=550$ MeV versus set (c) with $m_\sigma=600$ MeV, sets (a),(c) and
(d) with $\theta_P=-12^\circ$ versus set (b) with $\theta_P=-15^\circ$. The
scalar mixing angle is kept constant, $\theta_S=25^\circ$, in (a),(b),(c) and
increased to $\theta_S=27.5^\circ$ in set (d).}
\label{table-1}
\begin{tabular*}{\textwidth}{@{\extracolsep{\fill}}lrrrrrrrrl@{}}
\hline
     & \multicolumn{1}{c}{$m_\pi$}
     & \multicolumn{1}{c}{$m_K$}
     & \multicolumn{1}{c}{$m_\eta$}
     & \multicolumn{1}{c}{$m_{\eta'}$}
     & \multicolumn{1}{c}{$f_\pi$}
     & \multicolumn{1}{c}{$f_K$}
     & \multicolumn{1}{c}{$m_\kappa$}
     & \multicolumn{1}{c}{$m_{a_0}$}
     & \multicolumn{1}{c}{$m_{f_0}$}
\\
\hline
     & 138* & 494* & 547* & 958*  & 92*  & 113* & 850* &980* & 980*   \\
\hline
\end{tabular*}
\end{table*}

\begin{table*}
\caption{Parameter sets of the model: $\hat m, m_s$, and $\Lambda$ are given
         in MeV. The couplings have the following units: $[G]=$ GeV$^{-2}$,
         $[\kappa ]=$ GeV$^{-5}$, $[g_1]=[g_2]=$ GeV$^{-8}$. We also show here
         the values of constituent quark masses $\hat M$ and $M_s$ in MeV.
         See also caption of Table \ref{table-1}.}
\label{table-2}
\begin{tabular*}{\textwidth}{@{\extracolsep{\fill}}lrrrrrrrrl@{}}
\hline
Sets & \multicolumn{1}{c}{$\hat m$}
     & \multicolumn{1}{c}{$m_s$}
     & \multicolumn{1}{c}{$\hat M$}
     & \multicolumn{1}{c}{$M_s$}
     & \multicolumn{1}{c}{$\Lambda$}
     & \multicolumn{1}{c}{$G$}
     & \multicolumn{1}{c}{$-\kappa$}
     & \multicolumn{1}{c}{$g_1$}
     & \multicolumn{1}{c}{$g_2$} \\
\hline
a  & 4.0* & 100* & 372 & 541 & 830  & 9.74  & 121.1   & 3136  &133  \\
b  & 4.0* & 100* & 372 & 542 & 829  & 9.83  & 118.5   & 3305  &-158 \\
c  & 4.0* & 100* & 370 & 539 & 830  &10.45  & 120.3   & 2081  &102  \\
d  & 4.0* & 100* & 373 & 544 & 828  &10.48  & 122.0   & 3284  &173  \\
\hline
\end{tabular*}
\end{table*}

\begin{table*}
\caption{Explicit symmetry breaking interaction couplings. The couplings have
the following units: $[\kappa_1]=$ GeV$^{-1}$, $[\kappa_2]=$ GeV$^{-3}$,
$[g_3]=[g_4]=$ GeV$^{-6}$, $[g_5]=[g_6]=[g_7]=[g_8]=$ GeV$^{-4}$,
$[g_9]=[g_{10}]=$ GeV$^{-2}$. See also caption of Table \ref{table-1}.}
\label{table-3}
\begin{tabular*}{\textwidth}{@{\extracolsep{\fill}}lrrrrrrrrrl@{}}
\hline
Sets  & \multicolumn{1}{c}{$\kappa_1$}
      & \multicolumn{1}{c}{$\kappa_2$}
      & \multicolumn{1}{c}{$-g_3$}
      & \multicolumn{1}{c}{$g_4$}
      & \multicolumn{1}{c}{$g_5$}
      & \multicolumn{1}{c}{$-g_6$}
      & \multicolumn{1}{c}{$-g_7$}
      & \multicolumn{1}{c}{$g_8$}
      & \multicolumn{1}{c}{$g_9$}
      & \multicolumn{1}{c}{$g_{10}$} \\
\hline
a  &0* & 6.14  & 6338 & 657 & 210 & 1618  & 105  & -65  &0* &0*  \\
b  &0* & 5.61  & 6472 & 702 & 210 & 1668  & 100  & -38  &0* &0*  \\
c  &0* & 6.12  & 6214 & 464 & 207 & 1598  & 133  & -66  &0* &0*  \\
d  &0* & 6.17  & 6497 & 1235& 213 & 1642  & 13.3 & -64  &0* &0*  \\
\hline
\end{tabular*}
\end{table*}

\begin{table*}
\caption{Strong decays of the scalar mesons, $m_R$ is the resonance mass in
MeV, $\Gamma^{BW}$ and $\Gamma^{Fl}$ are the Breit-Wigner width and the Flatt\'e
distribution width in GeV, $R^S=\frac{{\bar g}^S_K}{{\bar g}_\beta}$.}
\label{table-4}
\begin{tabular*}{\textwidth}{@{\extracolsep{\fill}}lrrrrrrrrrl@{}}
\hline
Set   & \multicolumn{1}{c}{Decays}
      & \multicolumn{1}{c}{$m_R$}
      & \multicolumn{1}{c}{$\Gamma^{BW}$}
      & \multicolumn{1}{c}{$\Gamma^{Fl}$}
      & \multicolumn{1}{c}{${\bar g}_\beta$}
      & \multicolumn{1}{c}{${\bar g}^S_K$}
      & \multicolumn{1}{c}{$R^S$}
      & \multicolumn{1}{c}{$\theta_{P}$}
      & \multicolumn{1}{c}{$\theta_{S}$}
\\
\hline
a     &$\sigma\to\pi\pi$  &550  &465 &   & 1.95  &0.97  & 0.497 & -12 & 25 \\
      &$f_0\to\pi\pi$     &980  &108 &60 & 0.23  &0.32  & 1.397     \\
      &$\kappa\to K\pi$   &850  &310 &   & 1.2   & 0    &           \\
      &$a_0\to\eta\pi$    &980  &419 & 45& 1.32  &2.69  & 2.05       \\
\hline
Set   & \multicolumn{1}{c}{Decays}
      & \multicolumn{1}{c}{$m_R$}
      & \multicolumn{1}{c}{$\Gamma^{BW}$}
      & \multicolumn{1}{c}{$\Gamma^{Fl}$}
      & \multicolumn{1}{c}{${\bar g}_\beta$}
      & \multicolumn{1}{c}{${\bar g}^S_K$}
      & \multicolumn{1}{c}{$R^S$}
      & \multicolumn{1}{c}{$\theta_{P}$}
      & \multicolumn{1}{c}{$\theta_{S}$}
\\
\hline
b     &$\sigma\to\pi\pi$ &550 &465 &    & 1.955  &0.986  & 0.504 & -15& 25 \\
      &$f_0\to\pi\pi$    &980 &108 & 60 & 0.230  &0.312  & 1.356  \\
      &$\kappa\to K\pi$  &850 &310 &    & 1.2    &0      &        \\
      &$a_0\to\eta\pi$   &980 &459 & 50 & 1.44   &2.805  & 1.944   \\
\hline
Set   & \multicolumn{1}{c}{Decays}
      & \multicolumn{1}{c}{$m_R$}
      & \multicolumn{1}{c}{$\Gamma^{BW}$}
      & \multicolumn{1}{c}{$\Gamma^{Fl}$}
      & \multicolumn{1}{c}{${\bar g}_\beta$}
      & \multicolumn{1}{c}{${\bar g}^S_K$}
      & \multicolumn{1}{c}{$R^S$}
      & \multicolumn{1}{c}{$\theta_{P}$}
      & \multicolumn{1}{c}{$\theta_{S}$}
\\
\hline
c     &$\sigma\to\pi\pi$ &600  &635 &    & 2.39 &1.52 & 0.61  & -12& 25 \\
      &$f_0\to\pi\pi$    &980  &108 & 61 & 0.23 &0.30 & 1.32   \\
      &$\kappa\to K\pi$  &850  &310 &    & 1.2  & 0   &        \\
      &$a_0\to \eta\pi$  &980  &419 & 46 & 1.31 &2.67 & 2.03   \\
\hline
Set   & \multicolumn{1}{c}{Decays}
      & \multicolumn{1}{c}{$m_R$}
      & \multicolumn{1}{c}{$\Gamma^{BW}$}
      & \multicolumn{1}{c}{$\Gamma^{Fl}$}
      & \multicolumn{1}{c}{${\bar g}_\beta$}
      & \multicolumn{1}{c}{${\bar g}^S_K$}
      & \multicolumn{1}{c}{$R^S$}
      & \multicolumn{1}{c}{$\theta_{P}$}
      & \multicolumn{1}{c}{$\theta_{S}$}
\\
\hline
d     &$\sigma\to\pi\pi$ &550  &461 &    & 1.94 &0.63 & 0.33  & -12& 27.5 \\
      &$f_0\to\pi\pi$    &980  &62  & 30 & 0.23 &0.30 & 3.90   \\
      &$\kappa\to K\pi$  &850  &310 &    & 1.2  & 0   &        \\
      &$a_0\to\eta\pi$   &980  &420 &46  & 1.32 &2.73 & 2.07   \\
\hline

\end{tabular*}
\end{table*}


\begin{table*}
\caption{Radiative decays of the scalar mesons $\Gamma_{S\gamma\gamma}$ in KeV ,
         $m_R$ is the resonance mass in MeV. }
\label{table-5}
\begin{tabular*}{\textwidth}{@{\extracolsep{\fill}}lrrrrrrrrrrrrrrrl@{}}
\hline
Set a & \multicolumn{1}{c}{$m_R$}
      & \multicolumn{1}{c}{$\Gamma_{S\gamma\gamma}$}
      & \multicolumn{1}{c}{$\mbox{Set b} $}
      & \multicolumn{1}{c}{$m_R$}
      & \multicolumn{1}{c}{$\Gamma_{S\gamma\gamma}$}
      & \multicolumn{1}{c}{$\mbox{Set c} $}
      & \multicolumn{1}{c}{$m_R$}
      & \multicolumn{1}{c}{$\Gamma_{S\gamma\gamma}$}
      & \multicolumn{1}{c}{$\mbox{Set d} $}
      & \multicolumn{1}{c}{$m_R$}
      & \multicolumn{1}{c}{$\Gamma_{S\gamma\gamma}$} \\
\hline
  $\sigma\rightarrow\gamma\gamma$ &550  &0.212
& $\sigma\rightarrow\gamma\gamma$ &550  &0.212
& $\sigma\rightarrow\gamma\gamma$ &600  &0.277
& $\sigma\rightarrow\gamma\gamma$ &550  &0.210 \\
  $f_0\rightarrow\gamma\gamma$    &980  &0.055
& $f_0\rightarrow\gamma\gamma$    &980  &0.055
& $f_0\rightarrow\gamma\gamma$    &980  &0.055
& $f_0\rightarrow\gamma\gamma$    &980  &0.080 \\
  $a_0\rightarrow \gamma\gamma$   &980  &0.389
& $a_0\rightarrow \gamma\gamma$   &980  &0.386
& $a_0\rightarrow \gamma\gamma$   &980  &0.392
& $a_0\rightarrow \gamma\gamma$   &980  &0.383 \\
\hline

\end{tabular*}
\end{table*}

\begin{table*}
\caption{Anomalous decays $\Gamma_{P\gamma\gamma}$ for sets (a) and (c) in KeV,
corresponding to $\theta_P=-12^\circ$, $m_R$ is the particle mass in MeV. [For
set (b), corresponding to $\theta_P=-15^\circ$, we have
$\Gamma_{\eta\gamma\gamma}=0.6$ KeV, $\Gamma_{\eta'\gamma\gamma}=4.8$ KeV.] }
\label{table-6}
\begin{tabular*}{\textwidth}{@{\extracolsep{\fill}}lrrrl@{}}
\hline
Decays    & \multicolumn{1}{c}{$m_R$}
          & \multicolumn{1}{c}{$\Gamma_{P\gamma\gamma}$}
          & \multicolumn{1}{c}{$\Gamma^{exp}_{P\gamma\gamma}$ \cite{PDT:2011}} \\
\hline
 $\pi^0\to\gamma\gamma$  &136  &0.00798 & $0.00774637\div 0.00810933$ \\
 $\eta\to\gamma\gamma$   &547  & 0.5239 & $(39.31\pm 0.2)\%\,\Gamma_{\mbox{tot}}
                                          =0.508\div 0.569 $ \\
 $\eta'\to\gamma\gamma$  &958  &5.225   & $(2.18\pm 0.08)\%\,\Gamma_{\mbox{tot}}
                                          =3.99\div 4.70$ \\
\hline
\end{tabular*}
\end{table*}

\begin{table*}
\caption{The coefficients $\mbox{coef}^{HK}$ and  $\mbox{coef}^{SPA}$ of the
heat kernel and of the SPA contributions to the total value of the coupling
$g_{SP_1P_2}$ resulting from the interaction Lagrangian for the open decay
channels. Values are for the neutral channels. Units are in GeV.}
\label{table-7}
\begin{tabular*}{\textwidth}{@{\extracolsep{\fill}}lrrrrrrrrrl@{}}
\hline
     $g_{SP_1P_2}$
             & \multicolumn{1}{c}{$\mbox{coef}^{HK}/g^3$}
             & \multicolumn{1}{c}{$\mbox{coef}^{SPA}/g^3$}
             & \multicolumn{1}{c}{$\mbox{total}/g^3$}
\\
\hline
$\sigma\pi^0\pi^0$   &-0.0450 & 0.0215 & -0.0235    \\
$f_0\pi^0\pi^0$      &-0.0061 &-0.0047 & -0.0109     \\
$\kappa^0 {\bar K}^0\pi^0$ &0.0660 &-0.0257 & 0.0403 \\
$a_0^0\eta\pi^0$     &-0.0666 &-0.0178 & -0.0844 \\
\hline
\end{tabular*}
\end{table*}

\begin{table*}
\caption{The coefficients $\mbox{coef}^{HK}$ and  $\mbox{coef}^{SPA}$ of the
heat kernel and of the SPA  contributions to the total value of the coupling
$g_{SK\bar K}$ resulting from the interaction Lagrangian. Values are for the
neutral channels. Units are in GeV.}
\label{table-8}
\begin{tabular*}{\textwidth}{@{\extracolsep{\fill}}lrrrrrrrrrl@{}}
\hline
$g_{SK\bar KP_2}$
             & \multicolumn{1}{c}{$\mbox{coef}^{HK}/g^3$}
             & \multicolumn{1}{c}{$\mbox{coef}^{SPA}/g^3$}
             & \multicolumn{1}{c}{$\mbox{total}/g^3$}
\\
\hline
$\sigma K\bar K$  &-0.041  & 0.0178  &-0.0232 \\
$f_0 K \bar K$    & 0.118  &-0.081   & 0.0372 \\
$a_0^0 K\bar K$   & 0.0246 & 0.0968  & 0.121  \\
\hline
\end{tabular*}
\end{table*}


We obtain and understand the empirical mass assignment inside the light scalar
nonet as a consequence of the quark-mass dependent interactions, i.e. as the
result of some predominance of the explicit chiral symmetry breaking terms over
the dynamical chiral symmetry breaking ones for these states. Indeed, let us
consider the difference
\begin{eqnarray}
\label{a0-K}
   m_{a_0}^2-m_\kappa^2&\!=\!&2g^2\left(\frac{1}{H_{a_0}}-
   \frac{1}{H_\kappa}\right) \nonumber \\
                     &\!-\!&2(M_s+2\hat M)(M_s-\hat M).
\end{eqnarray}
The sign of this expression is a result of the competition of two terms. In the
chiral limit both of them are zero, since at $\hat\mu , \mu_s =0$ we obtain
$\hat M=M_s$ and $H_{a_0}=H_\kappa$, for $H_{a_0}$ and $H_\kappa$ being positive.
The splitting $H_\kappa >H_{a_0}$ is a necessary condition to get $m_{a_0}>
m_\kappa$. The following terms contribute to the difference
\begin{eqnarray}
\label{dif}
   H_\kappa -H_{a_0}
   &\!=\!&\kappa (h_s-\hat h)+2\kappa_2(\mu_s -\hat \mu ) \nonumber \\
   &\!-\!&g_2 (h_s^2+\hat hh_s -2\hat h^2) \nonumber \\
   &\!+\!&\frac{g_3}{2}\left(2\mu_s h_s + \mu_s\hat h + \hat\mu h_s
   -4\hat\mu\hat h\right) \nonumber \\
   &\!+\!&g_5\hat\mu (\mu_s -\hat\mu )+\frac{g_6}{2}
   \left(\mu_s^2-\hat\mu^2\right).
\end{eqnarray}
Accordingly, from this formula we deduce the ``anatomy'' of the
numerical fit, e.g. for set (d) (see next subsection):
\begin{eqnarray}
   m_{a_0}^2-m_\kappa^2&\!=\!&\left( [0.006]_\kappa + [0.046]_{\kappa_2}
                        + [6\times 10^{-4}]_{g_2} \right. \nonumber \\
                     &\!+\!&[0.938]_{g_3} + [0.003]_{g_5} +[-0.316]_{g_6}
                     \nonumber  \\
                     &\!-\!&\left. [0.44]_M = 0.24\right)\ \mbox{GeV}^2,
\end{eqnarray}
where the contributions of terms with corresponding coupling (see eq.
(\ref{dif})) are indicated in square brackets. The last number, marked by $M$,
is the value of the last term from (\ref{a0-K}). It is a contribution due to
the dynamical chiral symmetry breaking (in the presence of an explicit chiral
symmetry breaking). One can see that the $g_3$-interaction is the main reason
for the reverse ordering $m_{a_0}>m_\kappa$, the coupling $g_6$ being responsible
for the fine tuning of the result.

We now briefly comment on the role of parameters regarding the successful fit
of $f_\pi$ and $f_K$ as well as the ordering $m_K < m_\eta$. For these cases
many parameters are at work simultaneously. To illustrate this trend, we
deviate (arbitrarily) the values of $f_K$ and $m_\eta$ from their empirical
values, keeping the remaining observables fixed.

Let's consider first the weak decays. We take set (d) as reference and change
in the input data only $f_K=116$ MeV. As a result we obtain that the
constituent quark masses both decrease to ${\hat M} =351$ MeV and $M_s=533$
MeV, thus decreasing as well the normalization $g$ in order to fulfill eq.
(\ref{wd}). Regarding the interaction coupling strengths, the largest deviation
in absolute value is for $g_2$, which increases by $50\%$, followed by $g_1$
which decreases by $40\%$. The parameters $\{g_7,\kappa_2,g_3,g_4,g_6,\kappa\}$
decrease in the given order by $\{27,25,25,22,18,15\}$ parts in hundred, and
$g_8$ increases by $28\%$. The remaining parameters have much less significant
changes. We conclude that a very subtle interplay takes place involving
parameters related with and without the explicit symmetry breaking in this
case.

As for $m_K < m_\eta$: we take again set (d) as reference and change in the
input only the $\eta$ mass, lowering it to $\eta=490$ MeV. In this case the
largest changes are observed in $\{g_7,g_8,g_2\}$, with an increase of
$\{168,162,93\}$ per cent and a decrease in $\kappa_2$ by $73\%$, while a
lesser increase in $\{g_4,g_6,\kappa\}$ of $\{29,25,20\}$ and decrease of
$g_3$ by $16$ per cent is registered.

\subsection{Strong decays}

Let us now show the result of our global fitting of the model parameters. We
study the effect of having a slightly different $m_\sigma$ mass, sets (a), (b)
and (d) with $m_\sigma=550$ MeV versus set (c) with $m_\sigma=600$ MeV,  as well
as having  different pseudoscalar and scalar mixing angles, as described in the
caption of Table \ref{table-1}, with all other meson masses and weak decay
constants remaining fixed to the values there indicated.

Table \ref{table-2} contains the standard set of parameters, which are known
from previous considerations. Their values are not much affected by the quark
mass effects. We have already learned (as seen again in Table \ref{table-2})
that higher values of $g_1$ lead to the lower $\sigma$ mass
\cite{Osipov:2006a}. This eight-quark interaction violates Zweig's rule, since
it involves $q\bar q$ annihilation.

Table \ref{table-3} contains the couplings which are responsible for the
explicit chiral symmetry breaking effects in the interactions. Largest
variations are observed in the couplings $g_4$ and $g_7$ in set (d) as compared
to sets (a-c) and in $g_8$ between set (b) and the other sets. In the former
case it is related with the change of the scalar mixing angle and in the latter
with the change in the pseudoscalar mixing angle. The coupling $g_7$ is seen to
occur only in $(h_{ab}^{(1)})^{-1}$, thus it probes the mass spectrum of the
scalars, whereas $g_8$ appears only in $(h_{ab}^{(2)})^{-1}$, related to the mass
spectrum of the pseudoscalars. With all observables kept fixed, except the
mixing angle, changes in these couplings are obviously related to them.
Regarding $g_4$ it enters in both mass spectra. Comparing sets (a) and (c)
where both $\theta_S$ and $\theta_P$ are the same, but the $\sigma$ mass
different, show that that $g_4$ responds also to the change in the $\sigma$
mass.

The calculated values of quark condensates are approximately the same for all
sets: $-\langle\bar uu\rangle^{\frac{1}{3}}=232$\ MeV, and $-\langle\bar ss
\rangle^{\frac{1}{3}}=204$\ MeV. Our calculated values for the constituent quark
masses agree with the ones found in \cite{Georgi:1984,Weise:1990,Vogl:1990,Weise:1991}, showing their insensitivity to the new mass-dependent corrections.

In Table \ref{table-4} are shown the results for the strong decay widths of
the scalar mesons for the four different sets. The experimental status is as
follows. The  mass and width of the $\sigma$ meson quoted until recently had a
large uncertainty, $m_\sigma=(400\div 1200)$ MeV and a full width $\Gamma_\sigma
=(600\div 1000)$ MeV. Presently \cite{PDT:2011} it has been narrowed to
$m_\sigma=(400\div 550)$ MeV and $\Gamma_\sigma=(400\div 700)$ MeV. The result
based on the average over the dispersion analysis of \cite{Colangelo:2001,Caprini:2006,Kaminski:2011,Moussalam:2011} leads even to a very sharp value for the
pole position $M-i\Gamma/2=(446\pm 6)-(276\pm 5)$ MeV. The mass and full width
of the $f_0(980)$ meson are quoted as $m_{f_0(980)}=990\pm 20$ MeV and
$\Gamma_{f_0(980)}=40\div 100$ MeV and for the $a_0(980)$ meson as $m_{a_0(980)}
=980\pm 20$ MeV and $\Gamma_{a_0(980)}=50\div 100$ MeV. The results for the
$\kappa(800)$ quoted in the PDG table from a Breit-Wigner fit have the pole at
$(764\pm 63^{+71}_{-54})-i(306\pm 149^{+143}_{-82})$ MeV.

We obtain that the $\sigma$ mass and $\sigma\to\pi\pi$ decay are within the
recent limits for sets (a-b) and (d) while set (c) has a mass larger than the
upper limit by $\sim 50$ MeV. While in set (a-b) and (d) the calculated width
is smaller than the nominal mass of the resonance, the opposite behavior is
seen in set (c). The coupling strength ${\bar g}_{\sigma\pi\pi}$ increases
comparing e.g. set(a) to (c) explaining the larger width, however the ratio
$R^\sigma=\frac{{\bar g}^\sigma_K}{{\bar g}_{\sigma\pi\pi}}$ of the $\sigma$ to
kaon and to the pion couplings also increases by $20\%$. The obtained ratios
for $R^\sigma$ are in agreement with the experimental value $R^\sigma_{exp}=0.5
\pm 0.1$ in \cite{Bugg:2006} for sets (a-c) and slightly below for set (d). We
expect some effect on the width if these channels were taken into account, but
only a moderate one since the coupling to pions dominates, $R^{\sigma}\sim 0.3
\div 0.5$.

The decay width for $\kappa(800)\to K\pi\sim 310$ MeV is smaller roughly by a
factor two than the quoted central value but lies still within the limits. The
ratio of the couplings $\frac{{\bar g}_{\kappa K\pi}}{{\bar g}_{\sigma\pi\pi}}\frac{m_\kappa^2}{m_\sigma^2}=1.5$ (the ratio of meson masses corrects for the
different definitions of the couplings in \cite{Bugg:2006}) is within the
experimental values in \cite{Bugg:2006}, as opposed to the $q\bar q$ and $q^2
{\bar q}^2$ model approaches considered in the same paper.

The widths of the $a_0(980)\to\pi\eta$ and $f_0(980)\to\pi\pi$ decays are well
accomodated within a Flatt\'e description. We read the width at half maximum of
the elastic cross section in Figs. 1 and 2, respectively Note the huge
reduction in width in the case of the $a_0(980)$ meson when the kaon channels
are taken into account. This possibility was already noticed by Flatt\'e in
his analysis \cite{Flatte:1976}. This is explained in our description by the
ratio $R^{a_0}\sim 2$ showing the dominant component to be in the coupling to
the kaons.

As demonstrated in \cite{Baru:2005} the ratio
$R^S=\frac{{\bar g}^S_K}{{\bar g}^\beta}$ of the couplings is a relatively
stable quantity in despite of the large fluctuations in the experimental values
extracted for the individual couplings. Our calculated $R^S$ are compatible
with the indicated values in \cite{Baru:2005}. It should be emphasized that
the ratio $R^{f_0}=\frac{{\bar g}^{f_0}_K}{{\bar g}_{f_0\pi\pi}}$ is strongly
dependent on the mixing angle $\theta_S$ of the scalar sector. As can be seen
comparing sets (a-c) with set (d) the increase in $\theta_S$ is responsible for
the larger ratio $R^{f_0}=3.9$ in set (d), which agrees well with the
experimental value $R^{f_0}_{exp}=4.21\pm 0.46$ of BES \cite{BES}. An often
considered quantity is the crossed ratio $r=\frac{R^{f_0}}{R^{a_0}}$, usually
assumed to be larger than unity. The $a_0(980)$ does not depend on the
$\theta_S$ mixing angle (an eventual correlation with the $f_0(980)$ meson
through isospin mixing is discarded here), but does depend on the pseudoscalar
$\theta_P$ angle through its decay into the $\pi\eta$. The $\theta_P$ is fixed
in the pseudoscalar sector to yield the correct $\eta$ and $\eta'$ masses, as
well as their radiative two photon decay widths. Therefore the ratio
${R^{a_0}}$ of the $a_0$ couplings to kaons and to the $\pi\eta$ channels
remains approximately constant for all parameter sets $(R^{a_0})^{-1}\sim 0.5$.
This value is not too bad in comparison with the experimental quoted ratio
$(R^{a_0}_{exp})^{-1}=0.75\pm 0.11$ \cite{Hyams:1973}. Requiring the ratio $r>1$
constrains further the angle to be larger than $\theta_S\sim 26^\circ$.

On the other hand the ratio $R^{f_0}$ increases until $\theta_S$ reaches ideal
mixing. In the interval $\theta_{id}< \theta_S\le\frac{\pi}{4}$ it decreases
but stays much larger than the experimental accepted ratio, e.g. at $\theta_S=
44^\circ$ one has $R^{f_0}\sim 11$. The combined requirement $r>1$ and
$R^{f_0}_{exp}$ confines the mixing angle to the narrow window
$27^\circ<\theta_S<28^\circ$. From the point of view of the calculated strong
decay widths however the somewhat smaller angle $\theta_S=25^\circ$ is also
acceptable. Our interval of values for the mixing angle
$25^\circ<\theta_S<28^\circ$, corresponding to $-10.3^\circ < \bar\psi <-7.3^\circ$
are within the values $-14^\circ < \bar \psi <-3^\circ$ estimated in
\cite{Escribano:2002}, more specifically $\bar\psi \sim -9^\circ$ if a
Flatt\'e distribution is used in a complementarity approach of Chiral
Perturbation Theory and the Linear Sigma Model.

\subsection{Radiative decays}

The two photon decays of the pseudoscalars are in very good agreement with
data, (Table \ref{table-6}), the $\pi^0$ and $\eta$ in two photons are within
the experimental error bars, the $\eta'$ decay lies $10\%$ above the upper
limit for sets (a), (c) and (d), i.e. $\theta_P=-12^\circ$. In the case of set
(b), $\theta_P=-15^\circ$, the result for the $\eta'$ decay is at the upper
margin, and for the $\eta$ about $10\%$ above the upper boundary.

For the radiative widths of the $\sigma$, see Table \ref{table-5}, there is a
large spread in the experimental data from different facilities. Our results
for $\sigma\rightarrow\gamma\gamma$  only account for about $20\%$ of the
value $(1.2\pm 0.4)$ KeV \cite{Bernabeu:2008} obtained from the nucleon
electromagnetic polarizabilities, which is one of the lowest estimates for
this width. For the $f_0(980)\rightarrow\gamma\gamma$ the PDG average is quoted
as $(0.29^{+0.07}_{-0.06})$ KeV. Sets (a-c) yield approximately $20\%$ and set
(e) $30\%$ of this value. These results meet the current expectations that a
direct coupling to the photons via a quark loop are not sufficient to account
for the observed radiative widths of these mesons.

A natural question arises then why in our approach the strong widths can be
described reasonably well in all channels and the radiative ones fall short of
the empirical values for the $\sigma,f_0$ decays. This can be understood: only
the strong decays probe directly the multi-quark couplings $g_i$ contained in
the stationary phase (SPA) piece (\ref{lams}) of the total interaction
Lagrangian (\ref{lint}). Since this part of the Lagrangian has no derivative
terms only the heat kernel (HK) Lagrangian involves the electromagnetic
interaction, after minimal coupling. The information of the SPA conditions
which leaks through the gap equations to the electromagnetic sector is rather
weak; it is contained only in the wave function normalization which is the same
for all mesons, and the quark constituent masses and scale $\Lambda$ which
remain approximately constant in all parameter sets. Thus, effectively, the two
photon decays of the scalars yield a clean signature whether the
electromagnetic decay of the mesons proceeds dominantly through a $q\bar q$
channel or not.

This in turn ties up with the strength distribution in the HK and SPA
contributions to the coupling $g_{SPP}$  shown in Tables \ref{table-7} and
\ref{table-8} for set (d). The HK piece relates directly to the
meson-$q\bar q$ channel, the SPA part to the higher order multiquark
interactions.

Consider first the $a_0$ meson: the calculated $a_0(980)\to\gamma\gamma\sim
0.39$ KeV overestimates the present average PDG value $0.21^{+0.08}_{-0.04}$ and
points within our approach to the dominance of the direct one quark loop
coupling to photons of this meson.

This is corroborated by the fact that the large bare width that we obtain for
the $a_0\rightarrow\pi\eta$ decay is shown to stem mainly from the HK
coefficient represented with $80\%$ of the total strength, see Table
\ref{table-7}. The $a_0$ meson in the $q\bar q$ picture is composed only of
$u$ and $d$ quarks, thus its coupling to the $K\bar K$ mesons requires a
flavor change at the kaon vertices, as opposed to the $\eta\pi$ case. As can
be seen from a similar decomposition in HK and SPA contributions of the
$a_0 K\bar K$ coupling in Table \ref{table-8}, it is much more favorable to
couple to the kaons through the multiquark vertices, which now represent
$80\%$ of the total strength instead. Therefore for the overall strong decay
width it is important to take this mode into account through the two-channel
Flatt\'e distribution.  From the point of view of the two photon decay of
$a_0$, we note that a $\pi\eta$ loop does not couple directly to two photons
\footnote{In absence of the vector mesons. Their inclusion leads to this possibility through additional vertices $VP\gamma$, where V,P stand for vector and pseudosaclar mesons respectively \cite{Achasov:2010}.} and the decay proceeds
through the quark loop of u or d quarks with the large strength of the
corresponding HK component. To access the dominant SPA component the two photon
decay would have to proceed through coupling to the $K\bar K$ loop, a
sub-leading process in $N_c$ counting as compared to the direct $q\bar q$ loop.
Furthermore, due to the relatively large mass of the kaons, this loop is not
expected to contribute significantly.

Now let us analyze the $\sigma,f_0$ channels: there are substantial
contributions or cancellations from the SPA part. For the $f_0\pi\pi$ and
$f_0 K\bar K$ cases, one sees that the strength in the SPA coefficient is in
magnitude about $\frac{2}{3}$ of the HK coefficient for both cases, but changes
relative sign in the latter. In the $\sigma\pi\pi$ and $\sigma K\bar K$ cases,
the cancellations occur in both cases, with the SPA piece contributing about
half of the HK part. There is a subtle interplay about the HK and SPA
coefficients which finally add up to the correct description of the mass
spectra and strong decays of these mesons. The lack of a pronounced dominance
of the HK has as consequence that the $q\bar q$  coupling of these mesons to
the photons represents only a fraction of the total width. The remaining
strength must derive from the multiquark channels which should be included in
an extra step, taking into account explicitly meson loop contributions.

Regarding the strong decay of the $f_0$, one can further infer that because of
the stronger participation of the multi-quark interactions and because of
cancellations in the kaon channel as opposed to the pion channel, a coupling
to the kaon channel through the Flatt\'e approach is not imperative  to obtain
a reasonable magnitude of the width, as seen from the Table \ref{table-4}.

Rescattering effects have been shown in several approaches to yield the main
contribution, e.g. for the $\sigma\rightarrow\gamma \gamma$ extracted from the
dispersion analysis of $\gamma\gamma\rightarrow\pi^0\pi^0$ \cite{Oller:2008}.
Claims for a tetraquark structure \cite{Jaffe:1977} of the $\sigma$ meson were
forwarded e.g. in \cite{Giacosa:2006}, and in \cite{Achasov:2008} interpreted
as pion and kaon loop contributions. Our approach sheds light on these
phenomena from a different angle.

Finally we mention that the radiative decays of the scalar mesons have been
calculated a long time ago in a variant of the NJL model, with and without
meson loop contributions, \cite{Ebert:1997}. The amplitudes differ from ours
in two key aspects: we use the unified description for all non-anomalous
decays based on the generalized heat kernel approach which leads (i) to a
common wave function normalization for all mesons that implies the reduction
factor of $\sim \frac{2}{3}$ in the amplitude and in the case of the radiative
decays to (ii) the regularized one loop integrals carrying the factors
$(\frac{\Lambda^2}{\Lambda^2+M_i^2})^2$, in despite of the integrals being
finite. The latter reduces the amplitude by approximately half. The combined
effect is a dramatic reduction  by a factor $\sim 10$ in the decay widths, as
compared to \cite{Ebert:1997} for the quark loop contribution. Thus caution
must be used when it comes to interpret and comparing our numerical results
with seemingly related model calculations, e.g.
\cite{Volkov:2009},\cite{Schumacher:2011}.

Summarizing the results of sections IV B. and C., the strong decays calculated
from our tree level meson couplings encode leading and higher order $N_c$ and
multi-quark effects in combinations that account for the main bulk of the
empirical widths. The two photon decays of the scalars at leading order of the
bosonized Lagrangian yield complementary information, testing whether the
direct one quark loop coupling to photons is the dominant decay process. We
obtained that the $a_0$ meson decay into two photons proceeds mainly through
the $q\bar q$ loop, whereas for the $\sigma, f_0$ mesons we conclude that
higher order multi-quark interactions are necessary to account for the observed
widths. This does not mean that the $a_0$ meson is mainly a $q \bar q$ state,
but that the multi-quark component with the large strength in the two kaon
channel, important for the reduction of the $a_0 \pi\eta$ strong decay width,
is not the leading process in the two photon decay of this meson.

\begin{figure}[htb]
\hspace*{-3cm}
\includegraphics[height=6cm]{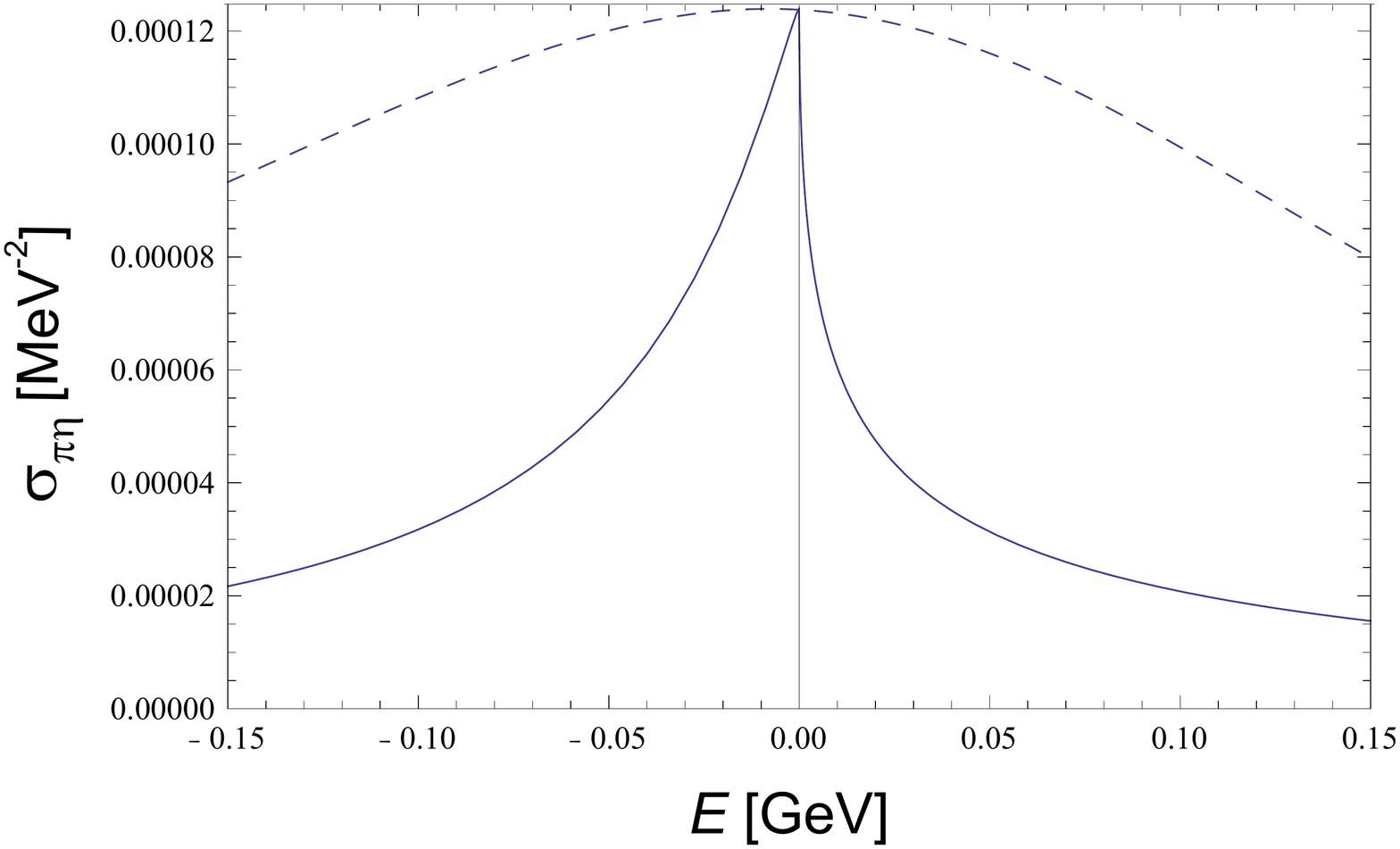}
\caption{\small The $\pi\eta$ cross section as function $E=\sqrt{s}-2m_K$ for
the $a_0$ resonance channel from the Flatt\'e distribution (solid line) with
parameters of set (b), ${\bar g}_{a_0\pi\eta}=1.44$, ${\bar g}^{a_0}_K=2.8$,
$R^{a_0}=1.944$. The width read at half peak value is $\Gamma^{Fl}=50$ MeV.
Dashed line corresponds to the single $\pi\eta$ channel.}

\end{figure}
\vspace{0.5cm}
\begin{figure}[htb]
\includegraphics[height=6cm]{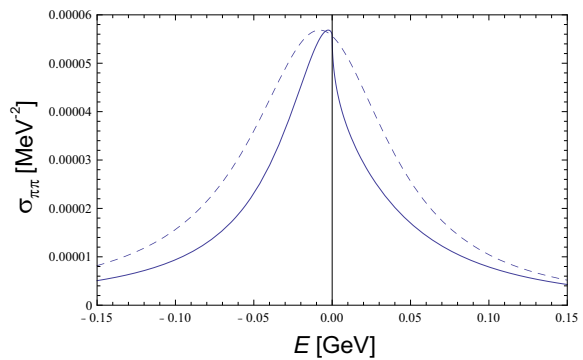}
\caption{\small The $\pi\pi$ cross section as function $E=\sqrt{s}-2m_K$ for
the $f_0$ resonance channel from the Flatt\'e distribution (solid line) with
parameters of set (b), ${\bar g}_{f_0\pi\pi}=0.23$, ${\bar g}^{f_0}_K=0.31$,
$R^{f_0}=1.36$. The width read at half peak value is $\Gamma^{Fl}=60$ MeV.
Dashed line corresponds to just the two pion channel.}
\end{figure}

\section{Concluding remarks}

In this paper we have generalized the effective multi-quark Lagrangians of the
NJL type by including higher order terms in the current quark-mass
expansion. The procedure is based on the very general assumption that the
scale of spontaneous chiral symmetry breaking determines the hierarchy of 
local multi-quark interactions. As a consequence, one can
distinguish a finite subset of vertices which are responsible for the explicit
chiral symmetry breaking at each order considered. We have classified these
vertices at next to leading order and studied the phenomenological consequences
of their inclusion in the Lagrangian.

We are led to a subset of ten quark-mass dependent interactions which enter the
Lagrangian at the same order as the 't Hooft determinant and eight quark terms
previously analyzed in the literature. From these, three are related with the
Manohar-Kaplan ambiguity, and the remaining seven with genuinely new vertices.
These new terms carry either signatures of violation of the Zweig-rule or of
admixtures of $q^2{\bar q}^2$ states to the quark-antiquark ones and are thus
potentially interesting candidates in the quest of analyzing the structure and
interaction dynamics of the low lying mesons.

We have derived the bosonized Lagrangian up to cubic order in the meson fields,
from which we obtain the meson spectra and their two body strong, weak and
electromagmetic decays. Here are our main conclusions:

(1) We fit the low lying pseudoscalar spectrum (the pseudo Goldstone $0^{-+}$
nonet) and weak decay constants of the pion and the kaon to perfect accuracy.
The fitting of the $\eta\!-\!\eta'$ mass splitting together with the overall
successful description of the whole set of low-energy pseudoscalar
characteristics is actually a solution for a long standing problem of NJL-type
models. We have found that the quark mass dependent interaction terms mainly
responsible for the fit belong to the class of OZI-violating interactions.
They represent additional corrections to the 't Hooft $U_A(1)$ breaking
mechanism. In the interaction terms independent of the quark masses, we
observe however that the $g_2$ coupling of the non OZI-violating $8q$
interactions carrying the signature of the $q^2{\bar q}^2$ states are also
relevant in fitting the $f_\pi, f_K$ values as well as for the ordering
$m_K < m_\eta$.

(2) We are also capable to describe the spectrum of the light scalar nonet. In
this case we identify the quark-mass interaction terms related with the four
quark admixtures to be the main source of the fit associated with the
$a_0(980)$ and $\kappa(800)$ meson masses. The primary term responsible for
the correct ordering carries interaction strength $g_3$, and some fine tuning
is due to the $g_6$ term.

(3) Regarding the mixing angle of the singlet-octet scalar states $\theta_S$ 
we have found that its value is particularly sensitive to the interaction term
proportional to $g_4$, which is OZI-violating. Together with the result that
the strength $g_1$ of the eight quark OZI-violating and quark mass independent
interaction term studied in earlier papers dictates the mass of the $\sigma
(500)$ meson, we conclude that these states are strongly affected by
OZI-violating short range forces.

(4) The calculation of the strong decays of the scalar mesons has revealed that
the present Lagrangian is capable of accounting for the decay widths within
the actual margins of empirical data. We corroborate other model calculations
in which the coupling of the $f_0(980)$ and $a_0(980)$ mesons to the $K \bar K$
channel is needed for the description of the decays $f_0(980)\rightarrow\pi\pi$
and $a_0(980)\rightarrow\pi\eta$. We find that this coupling is most crucial
for the latter process.

(5) The radiative decays of the scalar mesons into two photons show that the main
channel for the $a_0(980)$ decay proceeds through coupling to a quark-antiquark
state, while the radiative decays of singlet-octet states $\sigma, f_0$ must
proceed through more complex strutures. We refer to the full discussion given
in sections IV B and IV C.

(6) Finally, the radiative decays of the pseudoscalars are in very good
agreement with data.

\vspace{0.5cm}

{\bf Acknowledgements}

This work has been supported by the Funda\c{c}\~ao para a Ci\^encia e
Tecnologia, project: CERN/FP/116334/2010, developed under the iniciative QREN,
financed by UE/FEDER through COMPETE - Programa Operacional Factores de
Competitividade. This research is part of the EU Research Infrastructure
Integrating Activity Study of Strongly Interacting Matter (HadronPhysics3)
under the 7th Framework Programme of EU, Grant Agreement No. 283286.


\end{document}